\documentclass[twocolumn]{aastex701}
\usepackage{booktabs}
\usepackage{multirow}
\usepackage[utf8]{inputenc}

\begin{document}
% --------------------------------------------------------------
% -------------------------- COMMANDS --------------------------
% --------------------------------------------------------------
% This is a list of common commands that will be used, including
% lines, measurements, math, units, etc.

% Additionally, color definitions and commands color-coding
% text are also listed in here, now.  Co-authors, feel free to 
% add your own color command for your comments (if that's how
% you want to leave comments).
% --------------------------------------------------------------

% colors
\definecolor{notes}{HTML}{C70039}
\definecolor{emphasis_box}{HTML}{D1E1EC}
\definecolor{softgreen}{HTML}{468465}

% coloring text
\newcommand{\note}[1]{\textcolor{softgreen}{#1}}
\newcommand{\edits}[1]{\textcolor{notes}{#1}}

% code & models
\newcommand{\cloudy}{{\sc Cloudy}}
\newcommand{\bpass}{{\sc bpass}}

% box for emphasizing things
\newcommand{\colorme}[1]{\begin{tcolorbox}[width=\textwidth,
                            colback={emphasis_box}]
                            #1    
                         \end{tcolorbox}}

% LINES
\newcommand{\hi}{\hbox{\sc H\,i}}           % neutral HI gas
\newcommand{\lya}{\hbox{Ly$\alpha$}}        % Lya
\newcommand{\nv}{\hbox{\sc N\,v}}           % NV 1239,1243
\newcommand{\civ}{\hbox{\sc C\,iv}}         % CIV 1548,1551
\newcommand{\heii}{\hbox{He\,{\sc ii}}}     % HeII 1640
\newcommand{\oiiisemi}{\hbox{\sc O\,iii]}}  % OIII] 1661,1666
\newcommand{\ciii}{\hbox{\sc C\,iii]}}      % CIII] 1907,1909
\newcommand{\siiii}{\hbox{Si\,{\sc iii]}}}  % SiIII] 1883,1892
\newcommand{\mgii}{\hbox{Mg\,{\sc ii}}}     % MgII 2796,2803
\newcommand{\nev}{\hbox{[Ne\,{\sc v}]}}       % [NeV]
\newcommand{\oii}{\hbox{\sc [O\,ii]}}       % [OII]
\newcommand{\hd}{\hbox{\sc H$\delta$}}      % Hdelta 
\newcommand{\hb}{\hbox{\sc H$\beta$}}       % Hbeta 4863
\newcommand{\oiii}{\hbox{\sc [O\,iii]}}     % [OIII] 4959,5007
\newcommand{\ha}{\hbox{\sc H$\alpha$}}      % Ha 6563
\newcommand{\nii}{\hbox{[N\,{\sc ii}]}}     % [NII] 6583
\newcommand{\sii}{\hbox{[S\,{\sc ii}]}}     % [SII] 6717,6713
\newcommand{\siii}{\hbox{[S\,{\sc iii}]}}   % [SIII] 9069,9531
\newcommand{\pab}{\hbox{Pa$\beta$}}      % PaB 1.282micron
\newcommand{\hei}{\hbox{He\,{\sc i}}}
\newcommand{\pag}{\hbox{Pa$\gamma$}}
\newcommand{\neiii}{\hbox{[Ne\,{\sc iii]}}}  % [NeIII] 15.6micron
\newcommand{\siiiIR}{\hbox{[S\,{\sc iii]}}}  % [SIII] 18.7micron
\newcommand{\siv}{\hbox{[S\,{\sc iv]}}}  % [SIV] 10.5micron
\newcommand{\cii}{\hbox{[C\,{\sc ii}]}}     % [CII] 158micron
\newcommand{\het}{\hbox{\sc H$\eta$}}    % H eta 3890A

% Measurements from lines
\newcommand{\vlya}{\hbox{$\Delta$v$_{Ly\alpha}$}}
\newcommand{\vciv}{\hbox{$\Delta$v$_{CIV}$}}
\newcommand{\wciii}{\hbox{W$_{\text{\sc C\,iii}]}$}}
\newcommand{\wciv}{\hbox{W$_{\text{\sc C\,iv}}$}}
\newcommand{\wlya}{\hbox{W$_{\text{Ly$\alpha$}}$}}
\newcommand{\fesc}{\hbox{$f_{esc}$}}
\newcommand{\wline}{\hbox{W$_{line}$}}

% Units  (just a very big list)
\newcommand{\lam}{$\lambda$}
\newcommand{\unit}[1]{\ensuremath{\mathrm{\,#1}}\xspace}
\newcommand{\yr}{\unit{yr}}
\newcommand{\Gyr}{\unit{Gyr}}
\newcommand{\Myr}{\unit{Myr}}
\newcommand{\eV}{\unit{eV}}
\newcommand{\keV}{\unit{keV}}
\newcommand{\MeV}{\unit{MeV}}
\newcommand{\GeV}{\unit{GeV}}
\newcommand{\TeV}{\unit{TeV}}
\newcommand{\MB}{\unit{MB}}
\newcommand{\GB}{\unit{GB}}
\newcommand{\TB}{\unit{TB}}
\newcommand{\degree}{\ensuremath{{}^{\circ}}\xspace}
\newcommand{\degrees}{\degree}
\newcommand{\mas}{\unit{mas}}
\newcommand{\amin}{\unit{arcmin}}
\newcommand{\asec}{\unit{arcsec}}
\newcommand{\arcsecond}{$^{\prime\prime}$}
\newcommand{\angstrom}{\unit{\AA}}
\newcommand{\ang}{$\mbox{\AA}$}
\newcommand{\um}{\unit{$\mu$m}}
\newcommand{\cm}{\unit{cm}}
\newcommand{\km}{\unit{km}}
\newcommand{\kms}{\km \second^{-1}}
\newcommand{\pc}{\unit{pc}}
\newcommand{\kpc}{\unit{kpc}}
\newcommand{\second}{\unit{s}}
\newcommand{\us}{\unit{$\mu$s}}
\newcommand{\photons}{\unit{ph}}
\newcommand{\photon}{\unit{ph}}
\newcommand{\sr}{\unit{sr}}
\newcommand{\Msolar}{\unit{M_\odot}}
\newcommand{\Msun}{\unit{M_\odot}}
\newcommand{\Mstar}{\unit{M_{*}}}
\newcommand{\Zsolar}{\unit{Z_\odot}}
\newcommand{\Lsolar}{\unit{L_\odot}}
\newcommand{\Lsun}{\unit{L_\odot}}
\newcommand{\Lstar}{\unit{L_{*}}}
\newcommand{\Lum}{\ensuremath{ L }\xspace}
\newcommand{\Dsun}{\unit{D_\odot}}
\newcommand{\Dgc}{\ensuremath{D_{GC}}\xspace}
\newcommand{\Rgc}{\ensuremath{R_{GC}}\xspace}
\newcommand{\cmcubes}{\ensuremath{\cm^{3}\second^{-1}}\xspace}
\newcommand{\magn}{\unit{mag}}
\newcommand{\mmag}{\unit{mmag}}
\newcommand{\e}{\unit{e^{-}}}
\newcommand{\rms}{\unit{rms}}
\newcommand{\pix}{\unit{pix}}
\newcommand{\rmspix}{\unit{rms/pix}}
\newcommand{\ermspix}{\e \rmspix}
\newcommand{\Mv}{\ensuremath{M_{V}}\xspace}
\newcommand{\Muv}{\ensuremath{\text{M}_{\text{UV}}}\xspace}

% Telescopes
\newcommand{\hst}{{\it HST}}
\newcommand{\HST}{{\it HST}}
\newcommand{\jwst}{{\it JWST}}
\newcommand{\JWST}{{\it JWST}}
\newcommand{\spitzer}{{\it Spitzer}}
\newcommand{\Spitzer}{{\it Spitzer}}

% for references
% journals
%\newcommand{\apj}{ApJ}
%\newcommand{\apjl}{ApJL}
%\newcommand{\mnras}{MNRAS}
%\newcommand{\nat}{Nature}
%\newcommand{\aap}{A\&A}
%\newcommand{\aj}{AJ}

% misc 
\newcommand{\etal}{et al.~}

\title{Deep Spectroscopic Follow-Up of Maisie's Galaxy - A Typical Galaxy in the Early Universe}

\suppressAffiliations

\author[0000-0003-2366-8858]{Rebecca L. Larson}
\altaffiliation{Giacconi Postdoctoral Fellow}
\affiliation{Space Telescope Science Institute, 3700 San Martin Drive, Baltimore, MD 21218, USA}
\email[show]{rlarson@stsci.edu}  

\author[0000-0001-6251-4988]{Taylor A. Hutchison} 
\altaffiliation{NASA Postdoctoral Fellow}
\affiliation{Astrophysics Science Division, NASA Goddard Space Flight Center, 8800 Greenbelt Rd, Greenbelt, MD 20771, USA}
\email{taylor.hutchison@nasa.gov}

\author[0000-0001-8519-1130]{Steven L.\ Finkelstein}
\affiliation{Department of Astronomy, The University of Texas at Austin, Austin, TX 78712}
\affiliation{Cosmic Frontier Center, The University of Texas at Austin, Austin, TX 78712}
\email{stevenf@astro.as.utexas.edu}

\author[0000-0002-7959-8783]{Pablo Arrabal Haro}
\altaffiliation{NASA Postdoctoral Fellow}
\affiliation{Astrophysics Science Division, NASA Goddard Space Flight Center, 8800 Greenbelt Rd, Greenbelt, MD 20771, USA}
\email{pablo.arrabalharo@nasa.gov}

\author[0000-0001-7503-8482]{Casey Papovich}
\affiliation{Department of Physics and Astronomy, Texas A\&M University, College Station, TX, 77843-4242 USA}
\affiliation{George P. and Cynthia Woods Mitchell Institute for Fundamental Physics and Astronomy,\\ Texas A\&M University, College Station, TX, 77843-4242 USA}
\email{papovich@tamu.edu}

\author[0000-0003-3424-3230]{Weida Hu} 
\affiliation{Key Laboratory for Research in Galaxies and Cosmology, Shanghai Astronomical Observatory, Chinese Academy of Sciences, 80 Nandan Road, Shanghai 200030, People's Republic of China}
\email{weidahu@shao.ac.cn}

\author[0000-0002-7093-1877]{Javier \'Alvarez-M\'arquez}
\affiliation{Centro de Astrobiolog\'{\i}a (CAB), CSIC-INTA, Ctra. de Ajalvir km 4, Torrej\'on de Ardoz, E-28850, Madrid, Spain}
\email{jalvarez@cab.inta-csic.es}

\author[0000-0003-3987-0858]{Ruqiu Lin}
\affiliation{University of Massachusetts Amherst, 710 North Pleasant Street, Amherst, MA 01003-9305, USA}
\email{ruqiulin@umass.edu}

\author[0000-0002-7051-1100]{Jorge A. Zavala}
\affiliation{University of Massachusetts Amherst, 710 North Pleasant Street, Amherst, MA 01003-9305, USA}
\email{jzavala@umass.edu} 

\author[0000-0003-0212-2979]{Volker Bromm}
\affiliation{Department of Astronomy, The University of Texas at Austin, Austin, TX 78712}
\affiliation{Cosmic Frontier Center, The University of Texas at Austin, Austin, TX 78712}
\email{vbromm@astro.as.utexas.edu}

\author[0000-0001-7151-009X]{Nikko J.\ Cleri}
\affiliation{Department of Astronomy and Astrophysics, The Pennsylvania State University, University Park, PA 16802, USA}
\affiliation{Institute for Computational \& Data Sciences, The Pennsylvania State University, University Park, PA 16802, USA}
\affiliation{Institute for Gravitation and the Cosmos, The Pennsylvania State University, University Park, PA 16802, USA}
\email{cleri@psu.edu}

\author[0000-0002-5258-8761]{Abdurro'uf} 
\affiliation{Department of Astronomy, Indiana University, 727 East Third Street, Bloomington, IN 47405, USA}
\affiliation{Center for Astrophysical Sciences, Department of Physics and Astronomy, The Johns Hopkins University, 3400 N Charles St. Baltimore, MD 21218, USA}
\affiliation{Space Telescope Science Institute, 3700 San Martin Drive, Baltimore, MD 21218, USA}
\email{fnuabdur@iu.edu}

\author[0000-0002-8163-0172]{Brittany Vanderhoof}
\affil{Space Telescope Science Institute, 3700 San Martin Drive, Baltimore, MD 21218, USA}
\email{bvanderhoof@stsci.edu}

%%%%%%%%%% PLEASE ADD YOUR AUTHOR BLOCK BELOW %%%%%%%%%%%%%

\author[0000-0001-8534-7502]{Bren E. Backhaus}
\affil{Department of Physics and Astronomy, University of Kansas, Lawrence, KS 66045, USA}
\email{bren.backhaus@ku.edu}

\author[0000-0001-7410-7669]{Dan Coe}
\affiliation{Space Telescope Science Institute, 3700 San Martin Drive, Baltimore, MD 21218, USA}
\affiliation{Center for Astrophysical Sciences, Department of Physics and Astronomy, The Johns Hopkins University, 3400 N Charles St. Baltimore, MD 21218, USA}
\affiliation{Association of Universities for Research in Astronomy (AURA), Inc.~for the European Space Agency (ESA)}
\email{dcoe@stsci.edu}

\author[0000-0001-7113-2738]{Henry C. Ferguson}
\affiliation{Space Telescope Science Institute, 3700 San Martin Drive, Baltimore, MD 21218, USA}
\email{ferguson@stsci.edu}

\author[orcid=0009-0006-1252-206X]{Ananya Ganapathy}
\affiliation{Center for Astrophysical Sciences, Department of Physics and Astronomy, The Johns Hopkins University, 3400 N Charles St. Baltimore, MD 21218, USA}
\email{aganapa1@jh.edu}  

\author[0000-0001-9440-8872]{Norman A. Grogin}
\affiliation{Space Telescope Science Institute, 3700 San Martin Drive, Baltimore, MD 21218, USA}
\email{nagrogin@stsci.edu}

\author[0000-0002-3301-3321]{Michaela Hirschmann}
\affiliation{Institute of Physics, Lab for galaxy evolution and spectral modeling, Observatory of Sauverny, Chemin Pegasi 51, 1290 Versoix, Switzerland}
\email{michaela.hirschmann@epfl.ch}

\author[0000-0003-1187-4240]{Intae Jung}
\affiliation{Department of Astronomy and Space Science, Chungbuk National University, Cheongju, 28644, Republic of Korea}
\email{ijung@cbnu.ac.kr}

\author[0000-0001-9187-3605]{Jeyhan S. Kartaltepe}
\affiliation{Laboratory for Multiwavelength Astrophysics, School of Physics and Astronomy, Rochester Institute of Technology, 84 Lomb Memorial Drive, Rochester, NY 14623, USA}
\email{jeyhan@astro.rit.edu}

\author[0000-0002-6610-2048]{Anton M. Koekemoer}
\affiliation{Space Telescope Science Institute, 3700 San Martin Drive,
Baltimore, MD 21218, USA}
\email{koekemoer@stsci.edu}

\author[0000-0002-8360-3880]{Dale D. Kocevski}
\affiliation{Department of Physics and Astronomy, Colby College, Waterville, ME 04901, USA}
\email{dkocevski@colby.edu}

\author[0000-0003-1581-7825]{Ray A. Lucas}
\affiliation{Space Telescope Science Institute, 3700 San Martin Drive, Baltimore, MD 21218, USA}
\email{lucas@stsci.edu}

\author[0000-0003-4965-0402]{Alexa M.\ Morales}\altaffiliation{NSF Graduate Research Fellow}
\affiliation{Department of Astronomy, The University of Texas at Austin, Austin, TX 78712}
\affiliation{Cosmic Frontier Center, The University of Texas at Austin, Austin, TX 78712}
\email{alexa.morales@utexas.edu}

\author[0000-0000-0000-0000]{Pablo G. P\'erez-Gonz\'alez}
\affiliation{Centro de Astrobiolog\'{\i}a (CAB), CSIC-INTA, Ctra. de Ajalvir km 4, Torrej\'on de Ardoz, E-28850, Madrid, Spain}
\email{pgperez@cab.inta-csic.es}

\author[0000-0003-3382-5941]{Nor Pirzkal}
\affil{Space Telescope Science Institute, 3700 San Martin Drive, Baltimore, MD 21218, USA}
\email{npirzkal@stsci.edu}

\author[0000-0002-1410-0470]{Jonathan R. Trump}
\affiliation{Department of Physics, 196A Auditorium Road, Unit 3046, University of Connecticut, Storrs, CT 06269, USA}
\email{jonathan.trump@uconn.edu}

\author[0000-0003-3466-035X]{{L. Y. Aaron} {Yung}}
\altaffiliation{Giacconi Postdoctoral Fellow}
\affiliation{Space Telescope Science Institute, 3700 San Martin Drive, Baltimore, MD 21218, USA}
\email{yung@stsci.edu}

%%%%%%%%%%%%%%%%%%%%%%%%%%%%%%%%%%%%%%%%%%%%%%%%%%%%%%%%%%%

\collaboration{all}{The THRILS and C3PO Collaborations}

\shortauthors{Larson et al.}
\shorttitle{Spectroscopy of Maisie's Galaxy}

\begin{abstract}

The first several years of \jwst\ observations have yielded surprisingly large numbers of bright $z>10$ galaxies, with follow-up spectroscopy of many of these sources implying extreme star formation activity and/or AGN content. Here, we present a combination of two deep Cycle 3 NIRSpec G395M programs, totaling over 19 hours of exposure time, plus MIRI/LRS observations for one such high-redshift source: Maisie's Galaxy. We provide an updated redshift measurement of $z = 11.408 \pm 0.005$ for this source. Measurements of the \oii\ doublet in these data yield an electron density ($n_e = 108.56^{+873.9}_{-35.37}$) and a star-formation rate (SFR$_{\oii} = 1.3 \pm 0.35$), placing it along the star-formation main sequence (SFMS) and indicating that this is a much more typical, rather than extreme, source in the early Universe. We also report fluxes for the \oiii$\lambda$5008 and \neiii$\lambda$3869 lines that provide us with a $\log$(Ne3O2) $= -0.219 \pm 0.145$ and a $\log$(O32) $=0.724 \pm 0.191$. We estimate the metallicity ($Z/Z_{\odot} = 0.17 \pm 0.05$) and ionization parameter ($\log$(U) $= -2.26 \pm 0.13$) from the Ne3O2 ratio.  We place this galaxy in the context of other $z>10$ sources with similar line detections and compare the results to those obtained from SED fitting. The results suggest that we should go deeper with our observations to better understand the average galaxy population at these early times.  

\end{abstract}



%\input{commands}

\section{Introduction}

The study of galaxies in the early Universe was one of the key science drivers behind \jwst\ \citep[e.g.,][]{brommAR2011,robertsonAR2022}, and the telescope's first science results exceeded expectations \citep{adamoRev2025}. Many early \jwst/NIRCam \citep{gardner06,gardner23,rieke23} surveys yielded discoveries of galaxy candidates at $z>10$, those existing $<500$Myr after the Big Bang \citep[e.g.][]{castellano22, naidu22, finkelstein22, harikane22, donnan23, 2023ApJ...951L...1P, casey24, franco24}. Surprisingly, their number densities, especially at the bright end of the ultraviolet luminosity function (UVLF), were higher than those expected from most pre-\jwst\ models \citep{ArrabalHaro.2023, bouwens23, finkelstein23, finkelstein24}. These discoveries call for changes to our understanding of the physical conditions in early galaxies \citep{Boylan-Kolchin22,yung24}. 

Many scenarios have been proposed to explain this excess of bright galaxies so early on in cosmic history. These include such things as a top-heavy initial mass function (IMF; \citealt{trinca24, jeong_IMF2025,yung24}), lack of dust due to radiation-driven outflows \citep{ferrara23}, UV variability \citep{shen23,sun23}, a higher stellar-to-halo mass ratio and/or star-formation efficiency \citep{mason23, harikane23}, possible further boosting of the photon per stellar baryon yield via rapid rotation \citep{liu_CHE2025}, and the contribution of early active galactic nuclei (AGN; \citealt{cappelluti22}). To begin testing these theories, we must move beyond detections of these galaxy candidates to spectroscopic studies of their rest-frame UV and optical emission lines.

Already, a number of sources at $z>10$ have been spectroscopically confirmed and studied with \jwst/NIRSpec observations \citep[e.g.][]{curtislake22, bunker23, arrabalharo23, hsiao24a, castellano24, carniani24, Naidu.2026}. The rest-frame UV emission lines have strong diagnostic power but are often weak. \lya\ can constrain the source's ionization, and at these early times, detection of this feature implies that the source lies in already reionized regions of the Universe \citep{witstok25}. The carbon, nitrogen, and oxygen lines can be used to constrain chemical abundances, which tells us more about the interstellar medium (ISM) of these early galaxies. C and O are also predicted key drivers in the transition of SF mode from top-heavy to bottom-heavy (``critical metallicity", Z$_{crit}$). \citep[e.g.,][]{bromm_Zcrit2003}. However, to detect these powerful diagnostic lines, we have to go deep, and at these high redshifts, which requires long exposure times. 

Many of the $z>10$ galaxies that have been studied in detail with \jwst\ spectroscopy appear to be somewhat extreme systems. Several of these objects have shown emission-line features that are not typically found at lower redshifts. For example, GNz11 at $z=10.6$ shows evidence of a nitrogen abundance exceeding expectations given its low metallicity and young age \citep{bunker23, cameron23}. This could mean we are observing the formation of globular-cluster progenitors \citep{bekki23, dantona23, senchyna24, marqueschaves24, watanabe24}. Alternatively, a high C/O ratio in a galaxy at $z=12.5$, GSz12, may be the result of ejecta from a previous generation of Population III stars \citep{liu_CEMP2021,D'Eugenio.2024}. Additionally, a significant fraction of sources have shown evidence of AGN activity even at this early epoch \citep{goulding23, bogdan24, maiolino24,crespogomez26}, but they are under debate when you observe only the rest-frame optical lines \citep{Alvarez-Marquez.2025, Alvarez-Marquez.2026}. It is imperative that we understand these galaxies in detail and push our spectroscopic observations of $z>10$ sources to greater depths to determine whether these bright galaxies are typical of the galaxy population in the early Universe or represent extreme tip-of-the-iceberg sources.

\begin{figure*}[ht!]
    \centering
    \includegraphics[width=0.26\linewidth]{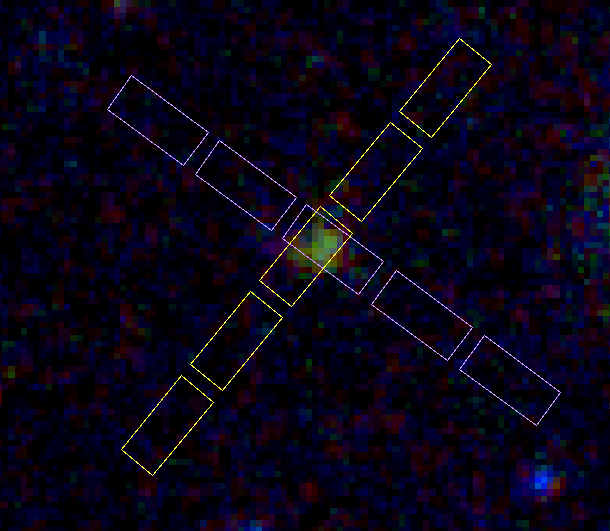}
    \includegraphics[width=0.72\linewidth,clip]{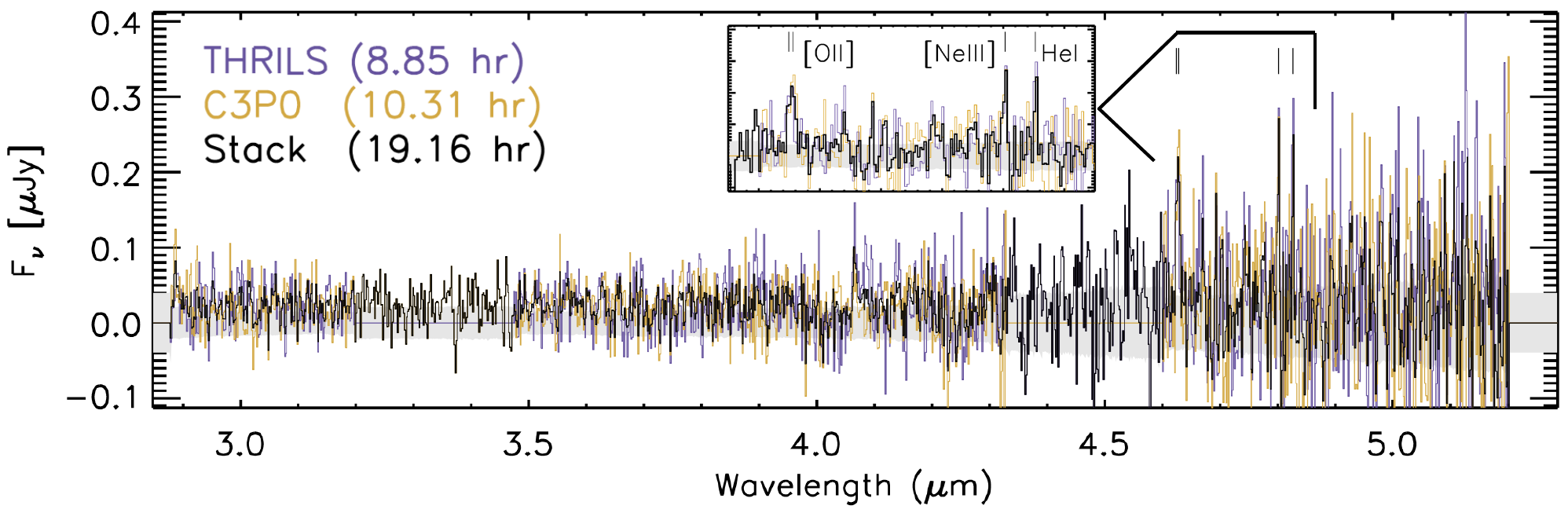}
    \caption{{\bf Left: } RGB color image of Maisie's galaxy using all seven NIRCam filters from the CEERS survey \citep{finkelstein25}. Overlaid are the NIRSpec shutters from THRILS (purple) and C3PO (gold). {\bf Right: }The G395M spectrum of Maisie's Galaxy from the THRILS (purple) and C3PO (gold) data, along with the combined stacked spectrum (black). The other program fills in the gaps in the detector that each program does not cover, since they do not overlap. The inset highlights the detected emission lines in the stacked spectrum.}% {\bf Bottom:} The stacked spectrum (black) compared to the DDT 2750 prism spectrum (green), where the \oii\ emission line was first detected. }
    \label{fig:StackedSpectrum}
\end{figure*}

Maisie's Galaxy (CEERS\_16943) was first discovered in the Cosmic Evolution Early Release Science (CEERS; ERS 1345, PI Finkelstein, \citealt{bagley22b, finkelstein25}) survey \jwst/NIRCam observations \citep{finkelstein22c}, and was one of the earliest \jwst\ $z >$ 10 discoveries \citep[e.g.][]{castellano22,naidu22,donnan22}.  This object was originally measured to have a photometric redshift ($z_{phot}$) of $\sim12$ \citep{finkelstein22c}. As the data pipelines and processing improved, the measurement was modified to $z_{phot}=11.5$ \citep{finkelstein23} and later updated to $z_{phot}=11.05 \pm 0.9$ \citep{arrabalharo23, harikane23}. 

Follow-up NIRSpec prism observations of this source (MPTID 1) and a few other key early discoveries were conducted through a Director's Discretionary Time program (DDT 2750; PI Arrabal Haro). They successfully detected the Lyman break and an \oii\ emission line feature, confirming the high-redshift nature and giving a spectroscopic redshift of $z=11.416\pm0.005$ \citep{arrabalharo23}, making this object the first early {\it JWST} discovery to be spectroscopically confirmed. However, this data was at the lowest available resolution with NIRSpec and did not allow detailed studies of the source's physical properties beyond a simple redshift confirmation. The spectrum was largely featureless, with only an \oii\ emission line detection and no rest-UV emission lines such as those found in several of the other early \jwst\ discoveries \citep[i.e.][]{bunker23,castellano24,witstok25}. This source, while one of the first $z>10$ sources discovered with \jwst, appears less extreme than many other discoveries in this era. 

In this paper, we present follow-up observations of this source with two \jwst/NIRSpec \citep{jakobsen22} Cycle 3 GO programs to obtain deep ($\sim 19$ hr) spectroscopy of Maisie's Galaxy at a higher spectral resolution. We also utilize \jwst/MIRI LRS \citep{rieke15,wright23} observations of this source. We describe the data in \S\ref{sec:data}, the line measurements and upper limits in \S\ref{sec:measurements}, update the spectro-photometric SED fitting measurements in \S\ref{sec:bagpipes}, and discuss the derived physical properties of the source and place it in context in \S\ref{sec:discussion}. We assume a flat $\Lambda$CDM cosmology, in line with the Planck results \citep{planck20}, of $H_0 = 70$\,km\,s$^{-1}$\,Mpc$^{-1}$, $\Omega_m = 0.3$, and $\Omega_b=0.05$ for all calculations in this paper.

\section{Data} \label{sec:data}

We have obtained deep NIRSpec data from two \jwst\ Cycle 3 GO programs of Maisie's Galaxy \citep{finkelstein22c, finkelstein23, arrabalharo23}, a $z=11.4$ source at (214.9431460, 52.9424397) in the CANDELS/EGS field \citep{grogin11,koekemoer11,davis07}. Both programs use the multi-object spectroscopy (MOS) mode \citep{ferruit22}, the same MPTID for this source (102659), and have been reduced in the same way.
We also incorporate additional data from the \jwst/MIRI LRS \citep{wright23}, which covers the longer-wavelength range. 

% \begin{figure}
%     \centering
%     \includegraphics[width=0.95\linewidth]{MaisieShutterOverlay.png}
%     \caption{RGB color image of Maisie's galaxy using all seven NIRCam filters from the CEERS survey \citep{finkelstein25}. Overlaid are the NIRSpec shutters from THRILS (purple) and C3PO (gold). }
%     \label{fig:ShutterOverlay}
% \end{figure}

\subsection{THRILS Data}

The High-(Redshift$+$Ionization) Line Search (THRILS; GO \#5507, PIs T.\ Hutchison \& R.\ Larson; \citealt{hutchison25b}) program obtained data for this source in March of 2025. This program used the G395M grating in a 3-shutter nod pattern, with a total exposure time of 31862.136s (8.85 hours) on this source. See \citet{hutchison25b} for more details on this program's data. The shutter overlay on the image and the 1D spectrum from this dataset are shown in purple in Figure \ref{fig:StackedSpectrum}. The THRILS data used in this paper can be found in MAST: \dataset[10.17909/6q08-j867]{http://dx.doi.org/10.17909/6q08-j867}.

\subsection{C3PO Data}

The Carbon 3 Plus Oxygen (C3PO;\footnote{C3 named for both the restframe UV \ciii\ semi-forbidden emission and triply-ionized Carbon, C$^{3+}$} PIs C.\ Papovich, W.\ Hu, \& T.\ Hutchison, GO-5943; Papovich et al., in prep) obtained data in June of 2025 for this source at an almost-perpendicular position angle to the THRILS data (see Figure \ref{fig:StackedSpectrum}, left). They used both the G395M and G140M gratings in a 3-shutter nod pattern. For this source, their G395M data had a total exposure time of 3714.136s (10.31 hours), and their G140M data had a total exposure time of 47793.204s (13.28 hours). See Papovich et al. (in prep) for more details on the data from this program. The shutter overlay on the image and the G395M 1D spectrum from this dataset are shown in gold in Figure \ref{fig:StackedSpectrum}. The C3PO data used in this paper can be found in MAST: \dataset[10.17909/t83k-jc86]{http://dx.doi.org/10.17909/t83k-jc86}.

\subsection{MIRI LRS Data}
The MIRI spectroscopic observations were obtained as part of a Cycle 2 program (GO \#3703; PI: J. Zavala) using the Low Resolution Spectroscopy (LRS) mode. The program targeted two high-redshift candidates, GHz2 (data previously reported in \citealt{Calabro.2024,Zavala.2024,zavala25,ChavezOrtiz.2025}) and Maisie's Galaxy. The observations were carried out using the FASTR1 readout pattern and the following setup: three visits with 121 groups per integration, 16 integration per exposure, and 1 exposure per specification, with 2 dither positions ``along slit nod''. Each visit has an on-source time of 10828\,s, adding a total on-source time of  $\sim9\,$h. The MIRI data used in this paper can be found in MAST: \dataset[10.17909/n1yg-g608]{http://dx.doi.org/10.17909/n1yg-g608}.

\subsection{Data Reduction}
The data from both NIRSpec programs were reduced using \jwst\ Calibration Pipeline version 1.20.2 (\citealt{bushouse25}, DOI:10.5281/zenodo.17515973) with CRDS context jwst\_1464.pmap. The processing is based on standard pipeline parameters, with several modifications for specific steps. The target was treated as a point source, and no bar-shadow correction was applied. The default pipeline pathloss correction was employed, and the default reference files for the adopted CRDS context were used for the flux calibration.  To account for slit losses, we measure the flux of the source in a $0\farcs22$ circular aperture (which encompasses all the light) in the CEERS NIRCam/F444W science-background image. We then measure the flux within the rectangular NIRSpec shutter aperture, defined by our optimal extraction window. We find a $\sim30\%$ slitloss correction is required for both observations in order to match the NIRCam photometry, and apply this to the extracted 1D spectra.

The MIRI LRS observations were also reduced with the \jwst\ Calibration Pipeline version 1.20.2 and CRDS context jwst\_1464.pmap. We follow the data reduction presented in \cite{Alvarez-Marquez.2026}, which includes additional customized steps: wavelength masking, master and residual background subtraction, and sigma clipping to mitigate detector artifacts and cosmic-ray residuals. The final 1D extracted spectrum was produced using a $0\farcs22$ aperture and corrected for aperture losses using the standard JWST reference files (\texttt{jwst\_miri\_apcorr\_0017.fits}).

\subsection{Stacked G395M Spectrum}

The source is well-centered in both NIRSpec programs, minimizing systematic uncertainties when combining spectra. We map the two G395M spectra onto the C3PO wavelength array, as it is the deeper spectrum. We fill the detector gap from C3PO (gold) with wavelengths and data from THRILS (purple), and the THRILS gap with wavelengths and data from C3PO. We then perform an error-weighted stacking of the spectra, yielding a combined spectrum (black) at 38976.273s, or 19.16 hours depth (see Figure \ref{fig:StackedSpectrum}). The inset in this figure highlights the three significant emission line features: \oii$\lambda\lambda3727,3729$\AA, \neiii$\lambda 3869$\AA, and \hei$\lambda 3889$\AA,  described below. All subsequent measurements are all made on the stacked G395M spectrum.

\section{Measured Emission Lines}\label{sec:measurements}

Emission lines in the NIRSpec data were measured using an automated line-finding routine developed originally for \hst\ grism data \citep{larson18} and modified for \jwst/NIRSpec data \citep{larson23}. This code first searched the spectrum, fitting Gaussians to each wavelength pixel, and then identified any significant peaks. The strongest line (highest flux and largest SNR) is then used to measure the redshift of the spectroscopic redshift of the source based upon the brightest expected line in this filter at the photometric redshift. Once the redshift is measured, the code fits the expected emission lines using single- or double-Gaussian profiles, depending on the line type. Applying this code to our NIRSpec data, we find three significant ($>3\sigma$) emission-line features in the stacked spectrum: \hei$\lambda 3889$\AA, \neiii$\lambda 3869$\AA, and \oii$\lambda\lambda3727,3729$\AA, shown in Figures \ref{fig:MeasuredLines} and \ref{fig:oiilinefit}. Emission-line measurements are summarized in Table \ref{tab:linemeasurements}.

\subsection{\neiii}
\begin{figure}
    \centering
     \includegraphics[width=0.95\linewidth]{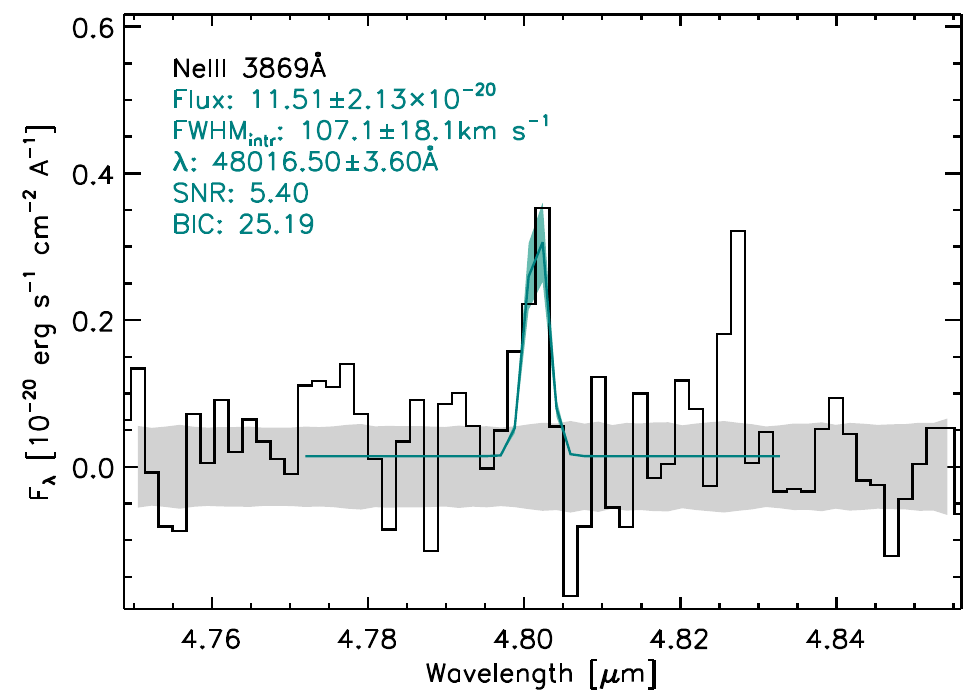}
      \includegraphics[width=0.95\linewidth]{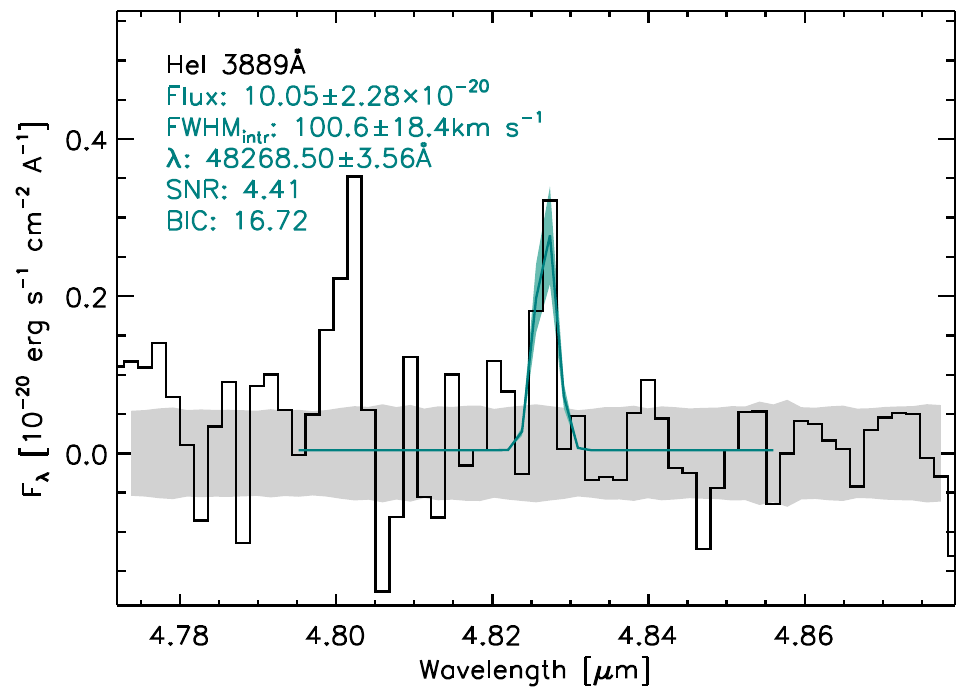}
    \caption{Fits to the \neiii\ (top) and \hei\ (bottom) emission lines in Maisie's Galaxy in our stacked 19.16-hour NIRSpec/G395M spectrum.}
    \label{fig:MeasuredLines}
\end{figure}

The strongest non-doublet line is the \neiii\ $\lambda 3869$\AA\  emission line at a signal-to-noise (SNR) of 5.40 with a measured flux of $11.51 \pm 2.13 \times 10^{-20}$ erg s$^{-1}$ cm$^{-2}$, and an intrinsic FWHM of $107.1\pm 18.1$ km s$^{-1}$, as shown in Figure \ref{fig:MeasuredLines} (top). We use this line to provide an updated redshift measurement for Maisie's Galaxy of $z=11.4082 \pm 0.0057$, in agreement with the redshift reported in \cite{arrabalharo23} from the \oii\ line detected in the prism spectrum at $z=11.416 \pm 0.005$. We use the intrinsic FWHM, $\pm 30$ km s$^{-1}$, of the \neiii\ line as the constraint for the fits to the other emission line features. 

\subsection{\hei}
The detected \hei\ $\lambda 3889$\AA\ line has a measured flux of $10.05 \pm 2.28 \times 10^{-20}$ erg s$^{-1}$ cm$^{-2}$, an intrinsic FWHM of $100.6\pm 18.4$ km s$^{-1}$, and a signal-to-noise (SNR) of 4.41, as shown in Figure \ref{fig:MeasuredLines} (bottom). We also note that this line is blended with the \het $\lambda 3890$\AA\ emission line feature at this resolution. The lack of detection of any other Balmer emission lines means we cannot directly measure the contribution of the \het\ line to the measured flux, but the upper limit on the \hd\ (see \S\ref{sec:upperlim}) line suggests that the \het\ contribution to this line flux is minimal. We therefore report this line flux as being that of the \hei\ line instead of the combination of the two lines. 

\begin{figure}[ht!]
    \centering
    \includegraphics[width=0.99\linewidth]{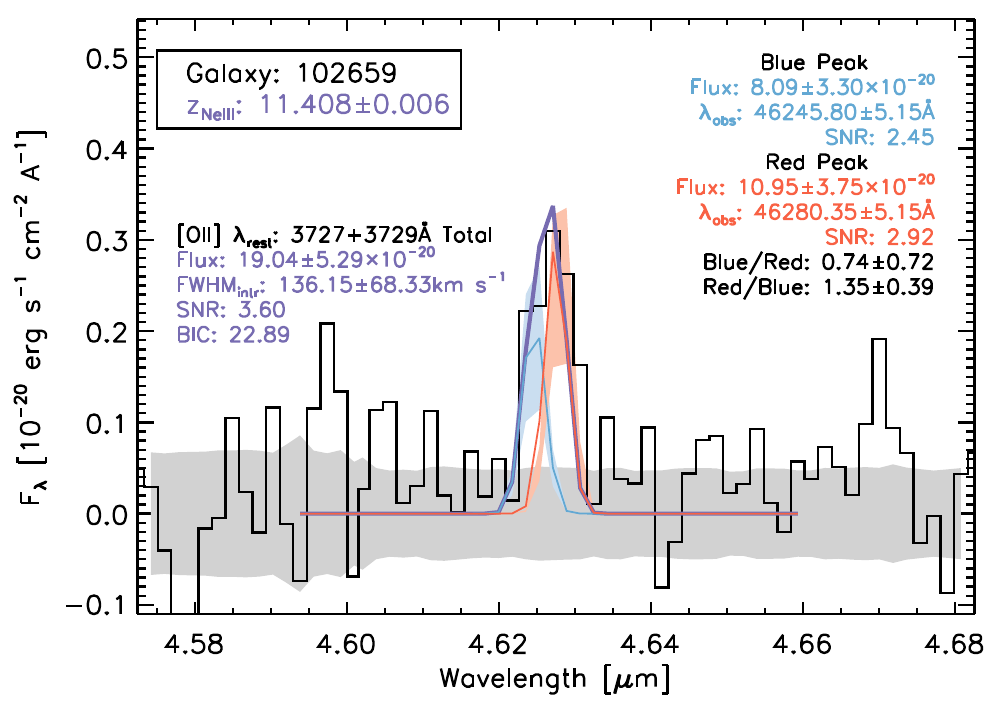}
    \caption{Fit to the \oii\ line doublet for Maisie's Galaxy. Fits to each of the doublet lines are shown in blue (\oii$\lambda 3727$) and red (\oii$\lambda 3729$), and the total fit to the combined doublet is shown in purple.}
    \label{fig:oiilinefit}
\end{figure}

\subsection{\oii}
The \oii\ emission feature is fit with two Gaussians, both restricted to within 30 km s$^{-1}$ of the intrinsic FWHM of the \neiii, as shown in Figure \ref{fig:oiilinefit}. Both emission lines thus have an individual intrinsic FWHM of $136.15 \pm 68.33$ km s$^{-1}$. We note that when only fitting a single Gaussian to this feature without restricting the FWHM, we find a best-fit observed FWHM of $\sim 450$ km s$^{-1}$, twice that of the observed \neiii\ FWHM, further indicating that we can resolve both of the \oii\ doublet emission lines. Each line is not formally detected individually with an SNR $< 3$, but the doublet fit has a combined SNR of 3.60. These lines are formally unresolved at this resolution ($R\lesssim 1400$). The doublet fit yields a line flux measurement for the \oii\ $\lambda 3727$\AA\ of $8.09 \pm 3.30 \times 10^{-20}$ erg s$^{-1}$ cm$^{-2}$, and for the \oii\ $\lambda 3729$\AA\ of $10.95 \pm 3.75 \times 10^{-20}$ erg s$^{-1}$ cm$^{-2}$. The total line flux for the \oii\ doublet is $19.04 \pm 5.29 \times 10^{-20}$ erg s$^{-1}$ cm$^{-2}$, less than that measured from the prism data, $28.5 \pm 6.7 \times 10^{-20}$ erg s$^{-1}$ cm$^{-2}$ \citep{arrabalharo23}, likely due to differences in the resolution, continuum fitting, slit losses, and data reduction between the data sets. 

The \oii\ flux ratio, $R_{\oii} = \oii \lambda 3729/3727$\AA, is measured as $1.35\pm0.39$. The full \oii\ flux is used in the \neiii$\lambda 3869$/\oii$\lambda$3727+3729\ (Ne3O2) ratio, which is found to be $0.605\pm0.138$ with $\log$(Ne3O2) $=-0.219\pm0.145$. These line ratios are reported in Table \ref{tab:linemeasurements}.

\subsection{\oiii}
The flux of the \oiii $\lambda \lambda 4960, 5008$ doublet is $134.50\pm45.87 \times 10^{-20} \rm erg\,s^{-1}\,cm^{-2}$.
The \oiii\ doublet is fit with two Gaussian profiles, which have fixed velocity dispersion and fixed redshifts for each line. We adopt a laboratory ratio of \oiii $\lambda5008$/\oiii $\lambda4960\ \sim 3$. 
Although \hb\ is blended with \oiii, we do not include \hb\ in the model, as it is not expected to be detected, given the lack of detected \ha, under the assumption of case B recombination and a dust-free model. 

With this additional line flux, we can calculate the O32 ratio: \oiii$\lambda 5008$/\oii$\lambda 3727,3729 = 5.30\pm2.33$. This gives $\log$(O32) $= 0.724\pm0.191$.
% with large errors given the low resolution of the MIRI/LRS data and the low signal-to-noise of the \oiii\ doublet.

\begin{figure}
    \centering
    \includegraphics[width=0.95\linewidth]{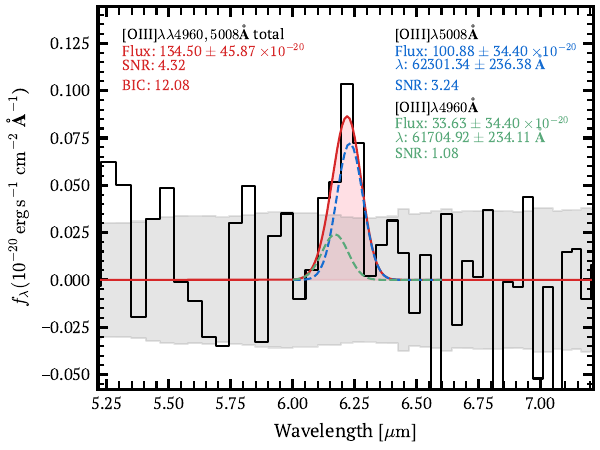}
    \caption{Fits to the \oiii\ line doublet (red) for Maisie's Galaxy. The \oiii$\lambda 5007$ (blue) and \oiii$\lambda 4959$ (green) lines are fixed with line width and a flux ratio of $\sim 3$.}
    \label{fig:oiiifit}
\end{figure}

\subsection{Upper Limits}\label{sec:upperlim}

Two other potentially strong emission-line features are covered by the wavelength range of this data but are not detected: \lya\ and \hd. Here, we set an upper limit by forcing a Gaussian fit at that wavelength and using its flux as the 1$\sigma$ error. We then calculate a $3\sigma$ upper limit for each emission line. For Maisie's Galaxy we find that \hd\ has an upper limit measurement of $13.20 \times 10^{-20}$ erg s$^{-1}$ cm$^{-2}$ from the $\sim19$ hours of stacked G395M spectra. This is consistent with the non-detection of \ha\ in the MIRI/LRS data. 

To measure the upper limit on \lya\, we first assume a zero velocity offset and fit a Gaussian at the systemic redshift. Using the $\sim13$ hour G140M spectrum from C3PO we find that \lya\ has an upper limit of $8.30 \times 10^{-20}$ erg s$^{-1}$ cm$^{-2}$. There are no line detections within 200 km s$^{-1}$ redward of the systemic redshift. These values are also reported in Table \ref{tab:linemeasurements}.

% \begin{figure*}
%     \centering
%     \includegraphics[width=0.45\linewidth]{Lya_UpperLimit.eps}
%     \includegraphics[width=0.45\linewidth]{Hd_UpperLimit.eps}
%     \caption{3$\sigma$ upper limits on \lya\ and \hd.}
%     \label{fig:UpperLimits}
% \end{figure*}

\begin{table}[ht!]
    \centering
    \begin{tabular}{ccc}
     Emission Line & Measured Value    \\
     \hline 
     $f_{\oiii Doublet}$ & 134.50$\pm$45.87 & \\
     $f_{\oiii \lambda5008}$ & 100.88$\pm$34.40 & \\
     $f_{\oiii \lambda4960}$ & 33.63$\pm$11.47 & \\
     $f_{\oii Doublet}$ & 19.04$\pm$5.29  \\
     $f_{\oii \lambda3727}$ & 8.09$\pm$3.30  \\
     $f_{\oii \lambda3729}$ & 10.95$\pm$3.75  \\
     $f_{\neiii \lambda3869}$ & 11.51$\pm$2.13 \\
     $f_{\hei \lambda 3889}$ & 10.05$\pm$2.28  \\
     \\
     Emission Line & Upper Limit \\
     \hline
     $f_{\lya \lambda 1216}$ & $<8.30$  \\
     $f_{\hd \lambda 4103}$ & $<13.20$  \\

     \\
     Ratio & Measured Value & \\
     \hline
     \oiii$\lambda 5008$/\oii$\lambda$3727,3729\ (O32) & 5.30$\pm$2.33 & \\
     $\log$(O32) & 0.724$\pm$0.191 & \\
     \neiii$\lambda 3869$/\oii$\lambda$3727,3729\ (Ne3O2) & $0.605\pm0.202$ & \\
     $\log$(Ne3O2) & $-0.219\pm0.145$ & \\     
     \oii$\lambda 3729$/$\lambda 3727$ (R$_{\oii}$) & $1.35\pm0.39$ & \\
     
     \\

    \end{tabular} 
    \caption{Measured line flux values and ratios for Maisie's Galaxy (MPTID 102659 - in both NIRSpec programs). All line fluxes are reported in $10^{-20}$ erg s$^{-1}$ cm$^{-2}$. \oiii\ values are measured from the MIRI/LRS spectrum. All other values are measured from the stacked G395M spectrum using both THRILS and C3PO, except for the upper limit on \lya\ which is measured from the C3PO G140M spectrum.}
    \label{tab:linemeasurements}
\end{table}

\section{Spectro-Photometric SED Fitting}\label{sec:bagpipes}

\begin{figure}[ht!]
    \centering
    \includegraphics[width=0.99\linewidth]{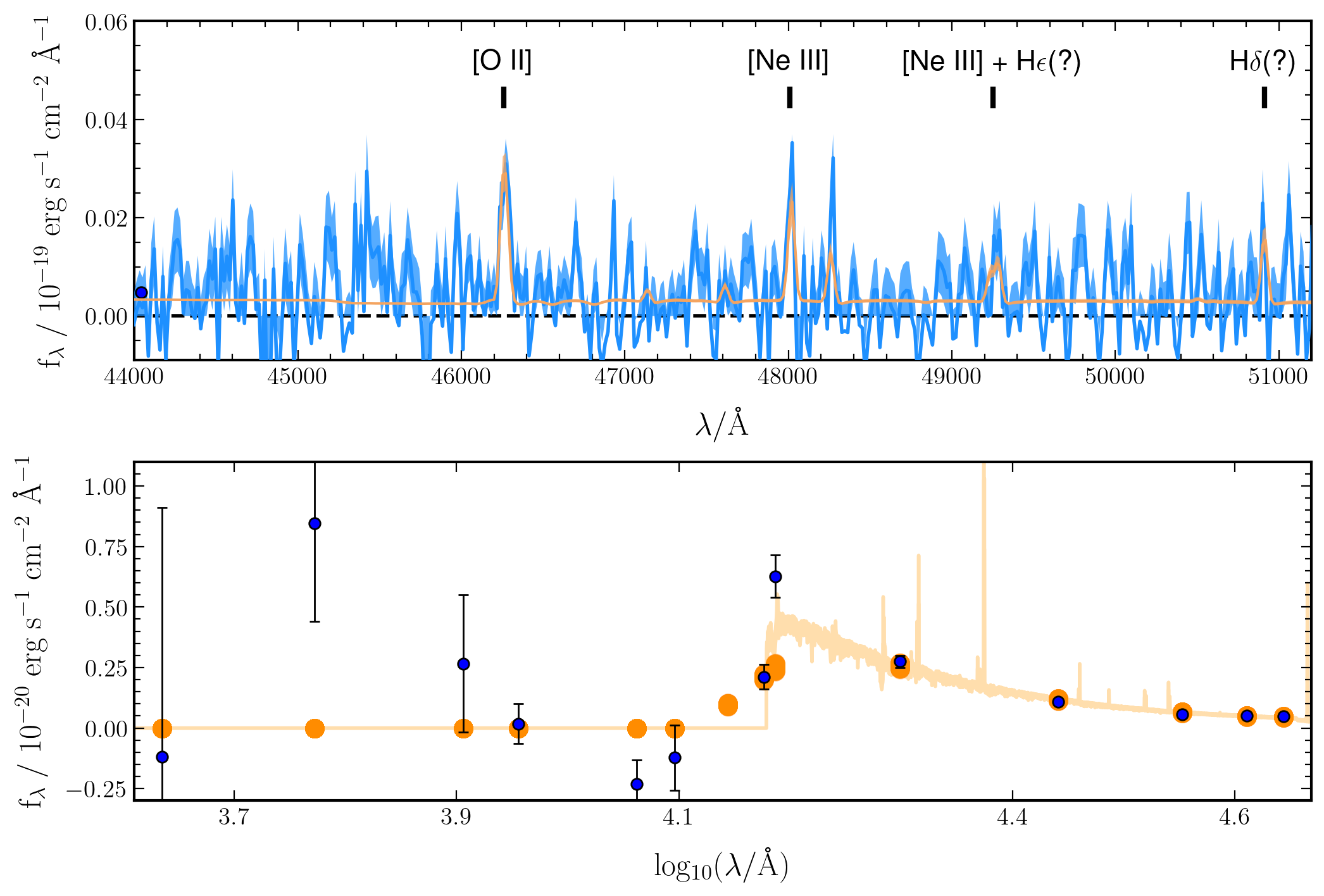}
    \caption{\textsc{BAGPIPES} \citep{carnall18, bagpipes} SED fit of Maisie's Galaxy using both the CEERS \citep{finkelstein25} photometry (bottom) and stacked G395M spectroscopy (top).}
    \label{fig:sed}
\end{figure}

We use the \textsc{BAGPIPES} (v1.3.2; \citealt{carnall18, bagpipes}) software to perform spectral energy distribution (SED) fitting for this source, incorporating both the photometry from \jwst/NIRCam and the spectroscopy from NIRSpec. Here, we use the C3PO G395M spectrum in conjunction with CEERS imaging \citep{finkelstein25} and the UNICORN photometric catalog (S. Finkelstein et al., in prep.), which emphasizes accurate colors and total fluxes. The combined spectrophotometric fits are shown in Figure \ref{fig:sed}.

% added by Papovich
The \textsc{BAGPIPES} fitting generally follows the method of \citet{papovich26}, which we briefly detail here.  We use the Binary Population and Spectral Synthesis (\textsc{BPASS}, v2.2.1) models from \citet{stanway18}, assuming a Chabrier IMF \citep{chabrier03} with an upper stellar mass cutoff of $300\,M_{\odot}$. We also adopt star formation histories (SFHs) parameterized by Gaussian processes \citep{iyer19}, as they offer the flexibility to model bursts and quenching events without imposing a prior on the widths of the star-formation time bins. We adopt the \citet{calzetti01} dust attenuation law, which is appropriate for more highly attenuated, blue star-forming galaxies \citep{salmon16}. We also assume that the stellar continua and nebular emission are attenuated by the same amount \citep{reddy15}, which is appropriate, as our galaxy is expected to have low dust attenuation and young stellar ages \citep{lecroq24}.

From this SED fitting, we measure a stellar mass for Maisie's Galaxy of $\log(M_*/M_{\odot}) = 8.73^{+0.14}_{-0.15}$. We infer a recent star-formation rate (in the last 10 Myr), $SFR_{10} = 3.99^{+0.49}_{-0.47}$ M$_{\odot}$ yr$^{-1}$ and a specific SFR of $\log sSFR_{10} = -8.02^{+0.18}_{-0.17}$ yr$^{-1}$. We also measure a dust attenuation, $A_v = 0.015_{-0.039}^{+0.011}$. These SED fitting results give a metallicity measurement of $Z = 0.129^{+0.083}_{-0.042}$ Z$_{\odot}$ and an ionization parameter of $\log U = -1.204^{+0.177}_{-0.176}$. These values can be found in Table \ref{tab:physicalproperties} as well as in Table \ref{tab:litsources}.

\section{Discussion}\label{sec:discussion}

\begin{table}[ht!]
    \centering
    \begin{tabular}{ccc}
 
     Quantity & Measured Value    & Units \\
     \hline 
     Redshift ($z$) & $11.4082\pm0.0057$ & \\ 
     SFR (\oii)$^\parallel$ & $1.30\pm0.36$ & M$_{\odot}$ yr$^{-1}$ \\
   % n$_e$ (T$_e$=10000K) & $119.69^{+719.01}_{-119.69}$ & cm$^{-3}$ \\
    % n$_e^*$  & $154.68^{+938.82}_{-154.68}$ & cm$^{-3}$ \\

   % n$_e$ (T$_e$=10000K) PyNeb & $159.21^{+1214.75}_{-82.75}$ & cm$^{-3}$ \\
     n$_e^*$  & $108.56^{+873.9}_{-35.37}$ & cm$^{-3}$ \\
     
  %   log(U) (Z=0.001) &$-2.65\pm0.09$ & \\ 
      
      Z$^\dagger$ [Ne3O2] & $0.17\pm0.05$ & Z$_{\odot}$ \\
       Z$^\dagger$ [O32]& $0.19\pm0.08$ & Z$_{\odot}$ \\
      %12 + log(O/H)$^\dagger$ & $7.92\pm0.07$ & \\
      log(U)$^\ddagger$ [Ne3O2] &$-2.26\pm0.13$ & \\ 
       log(U)$^\ddagger$ [O32] &$-2.38\pm0.$15 & \\
     %  log(U) (Z=0.008) &$-2.51\pm0.13$ & \\  
%       log(U) (Z=0.020) &$-2.07\pm0.22$ & \\ 
     \\
         Quantity & SED Inferred Value & Units \\
     \hline
     $\log (M_*)$ & $ 8.73^{+0.14}_{-0.15}$ & M$_{\odot}$ \\
     SFR$_{10}$ & $3.99^{+0.49}_{-0.47}$ & M$_{\odot}$ yr$^{-1}$ \\
     $\log (sSFR_{10})$ & $-8.02^{+0.18}_{-0.17}$ & yr$^{-1}$ \\
     A$_v$ & $0.015_{-0.011}^{+0.039}$ & \\
     Z & $0.13^{+0.08}_{-0.04}$ & Z$_{\odot}$ \\ 
     $\log (U) $ & $-1.204^{+0.177}_{-0.176}$ & \\

    \end{tabular} 
    \caption{Measured and inferred physical parameters for Maisie's Galaxy. Measured values are calculated using line fluxes and ratios and described in \S \ref{sec:discussion}. SED inferred values are from the \textsc{BAGPIPES} \citep{carnall18, bagpipes} fit to both the photometry and spectroscopy described in \S \ref{sec:bagpipes}. \\
    $^\parallel$ Using the equation from \citet{kewley04}. \\
    $^*$ Calculated using \textsc{PyNeb} \citep{pyneb} and assuming T$_e$=17000K as measured for the closest source in redshift with a direct measurement, MACS0647-JD1 \citep{hsiao24b, abdurrouf24}. \\
    $^\dagger$ Using the equation from \citet{sanders25c}. \\
    $^\ddagger$ Using the equation from \citet{witstok21}. 
    }
    \label{tab:physicalproperties}
\end{table}

\subsection{Lack of Ionized Bubble} \label{sec:ionizedbubble}

The non-detection of \lya\ implies that Maisie's Galaxy does not reside in an ionized bubble such that the \lya\ would resonantly scatter out of the surrounding neutral hydrogen in the intergalactic medium (IGM) and be detectable \citep[e.g.,][]{malhotra04,Smith_LAE2015}. This is different than JADES GSz13-1, a $z=13$ galaxy \citep{witstok25}, with detected significant \lya\ emission. The lack of other sources at this redshift in CEERS \citep{finkelstein23} indicates that it is not in an overdense region and therefore does not benefit from the combined ionizing power of multiple sources.

\subsection{Star-Formation Rate from \oii} \label{sec:sfroii}

\begin{figure}
    \centering
    \includegraphics[width=0.98\linewidth]{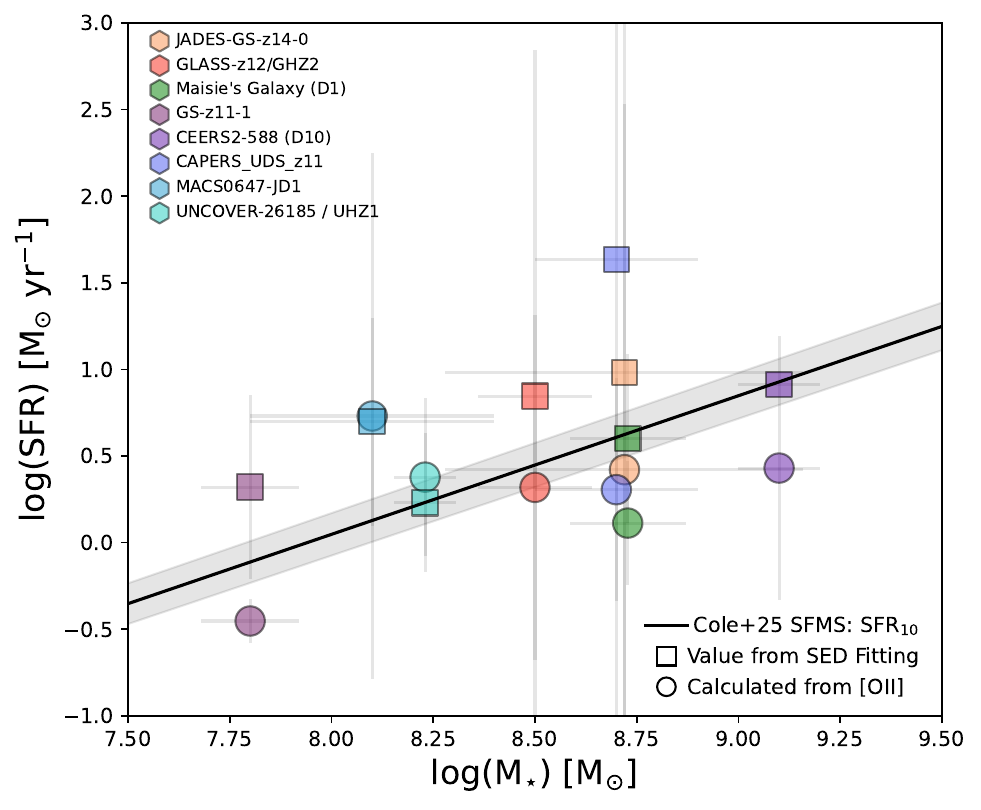}
    \caption{%\textbf{Left:} Specific star-formation rate (sSFR) vs redshift (z) for the sources in the literature at $z>10$ that have a measured \oii\ line flux or a sSFR reported. Maisie's galaxy is plotted in green. \textbf{Right:} 
    Star-formation rate (SFR) vs stellar mass (M$_{\star}$) for the sources in the literature at $z>10$ that have a measured \oii\ line flux (circles) and a SFR$_{10}$ from SED fitting (squares). The two measurements for Maisie's Galaxy are plotted in green. As discussed in \S\ref{sec:sfroii}, the SFR$_{\oii}$ can be considered a lower limit, and in almost all cases is smaller than that measured from SED fitting. The star-formation main sequence (SFMS) from \citet{cole25}, calculated from CEERS data using SFR$_{10}$ at $9 < z <12 $, is plotted as the black line. All of these sources fall along the SFMS line, within error.}
    \label{fig:zvssfr}
\end{figure}

Since these high redshifts push the \ha\ emission line feature out of the NIRSpec G395M coverage, the traditional feature for measuring star-formation rate (SFR) of a galaxy cannot be used without MIRI coverage. While we do have MIRI/LRS observations of Maisie's galaxy, the \ha\ line is not detected (Zavala et al. in prep). However, we can instead use the measured \oii\ emission line and compare it with the SED-fitting SFR measurement from \citet{arrabalharo23}, which is $2\pm1$ (M$_\odot$ yr$^{-1}$). \citet{vanderhoof22} find that if only \oii\ is available to measure the SFR, it is better to use the \citet{kewley04} relation rather than the \citet{kennicutt98} one, as the latter will overestimate the SFR. Our total $f_{\oii} = 19.04\pm5.29 \times 10^{-20}$ erg s$^{-1}$ cm$^{-2}$ which gives a $L_{\oii} = 3.49\pm 0.97 \times 10^{41}$ ergs$^{-1}$. Using equation 4 in \citet{kewley04} as given below 
$$ SFR_{\oii} [M_{\odot} yr^{-1}] = (3.72 \pm 0.93) \times 10^{-42} L_{\oii}[erg s^{-1}]$$
we get a SFR$_{\oii} = 1.30 \pm 0.36 $ M$_{\odot}$ yr$^{-1}$. 

We do not apply a dust correction to the \oii, as we do not consistently have this value for other sources in the literature. Given the A$_{v}$ = 0.015 from our SED fitting the correction is expected to be minor, but indicates that the SFR$_{\oii}$ is underestimated. While the \citet{kewley04} calculation does account for metallicity, the recent calibrations from \citet{zhuang19} do not go to the low metallicity we measure for this source (See \S\ref{sec:metallicity}). As you need a higher SFR to get a certain oxygen emission if the metallicity is low, this indicates that our SFR$_{\oii}$ is likely lower than the actual SFR. We find that our SFR$_{\oii}$ is consistent with the SFR measured from SED fitting by \cite{arrabalharo23}, who use the photometry plus the NIRSpec prism spectrum, but below the SFR in the last 10 Myr (SFR$_{10}$) of $3.99^{+0.49}_{-0.47}$, which we get from our \textsc{BAGPIPES} fit.

We also use this same method to calculate the SFR$_{\oii}$ for all of the sources in the literature at $z>10$ with measured \oii\ emission line fluxes \citep{helton25, castellano24, scholtz26, harikane26, kokorev25, bunker23, hsiao24a, napolitano25a, goulding23}. These SFR$_{\oii}$ values can be found in Table \ref{tab:measurements} and corresponding SFR$_{10}$ values from the literature are reported in Table \ref{tab:litsources}. We also find that the SFR$_{\oii}$ value is either equal to or lower than the SFR$_{10}$ reported from SED fitting in these other sources, further indicating that our SFR$_{\oii}$ values are underestimating the total SFR. This can be seen in Figure \ref{fig:zvssfr} where the SED-based SFR$_{10}$ values are plotted as squares in the same color as the corresponding SFR$_{\oii}$ values plotted as circles. When we plot these values versus stellar mass (M$_{\star}$), we find that Maisie's Galaxy and the other galaxies at this epoch fall along the star-formation main sequence (SFMS) as calculated by \citet{cole25} using photometric sources from the CEERS \citep{finkelstein25} survey at $ 9 <z<12$ (See Figure \ref{fig:zvssfr}).

%OII is can be affected by diffuse gas, if there's alpha enhancement the OII SFR could be higher than the SED value 

\begin{figure}[h!]
    \centering
    \includegraphics[width=0.95\linewidth]{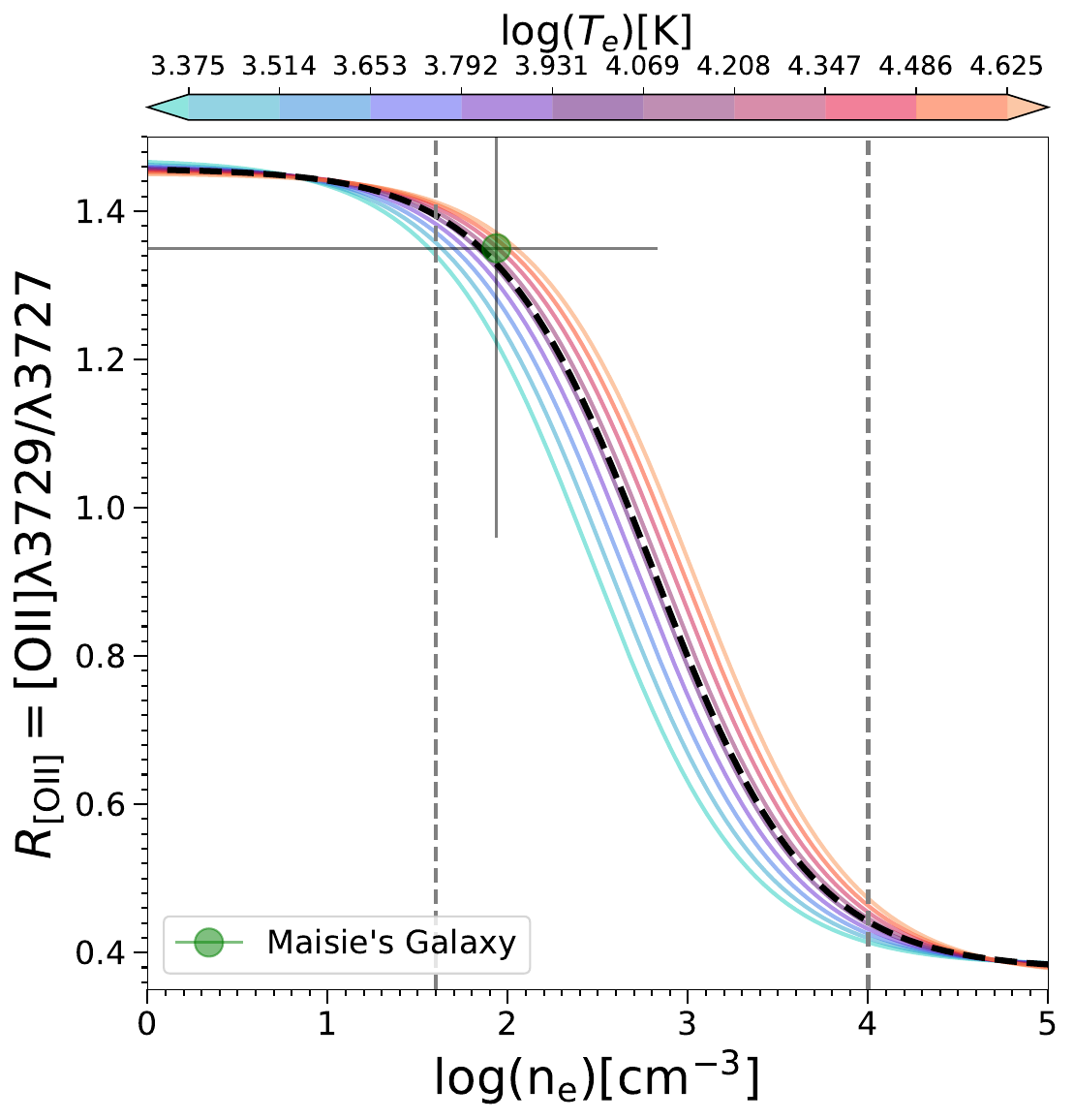}
    \caption{Plot of the electron density ($n_e$) vs \oii\ doublet ratio. The colored curves come from \citet{kewley19a} and the dashed line is the curve from \citet{sanders16}. Here we assume a $T_e = 17,000$ K, equal to that measured for the lensed $z=10.2$ galaxy in \citet{abdurrouf24, hsiao24b}. For our measured $R_{\oii} = 1.35\pm 0.39$ we get $n_e = 86.36^{+588.11}_{-86.36}$.}
    \label{fig:te}
\end{figure}

\subsection{Electron Density} \label{sec:density}

\begin{figure*}[ht!]
    \centering
    \includegraphics[width=0.95\linewidth]{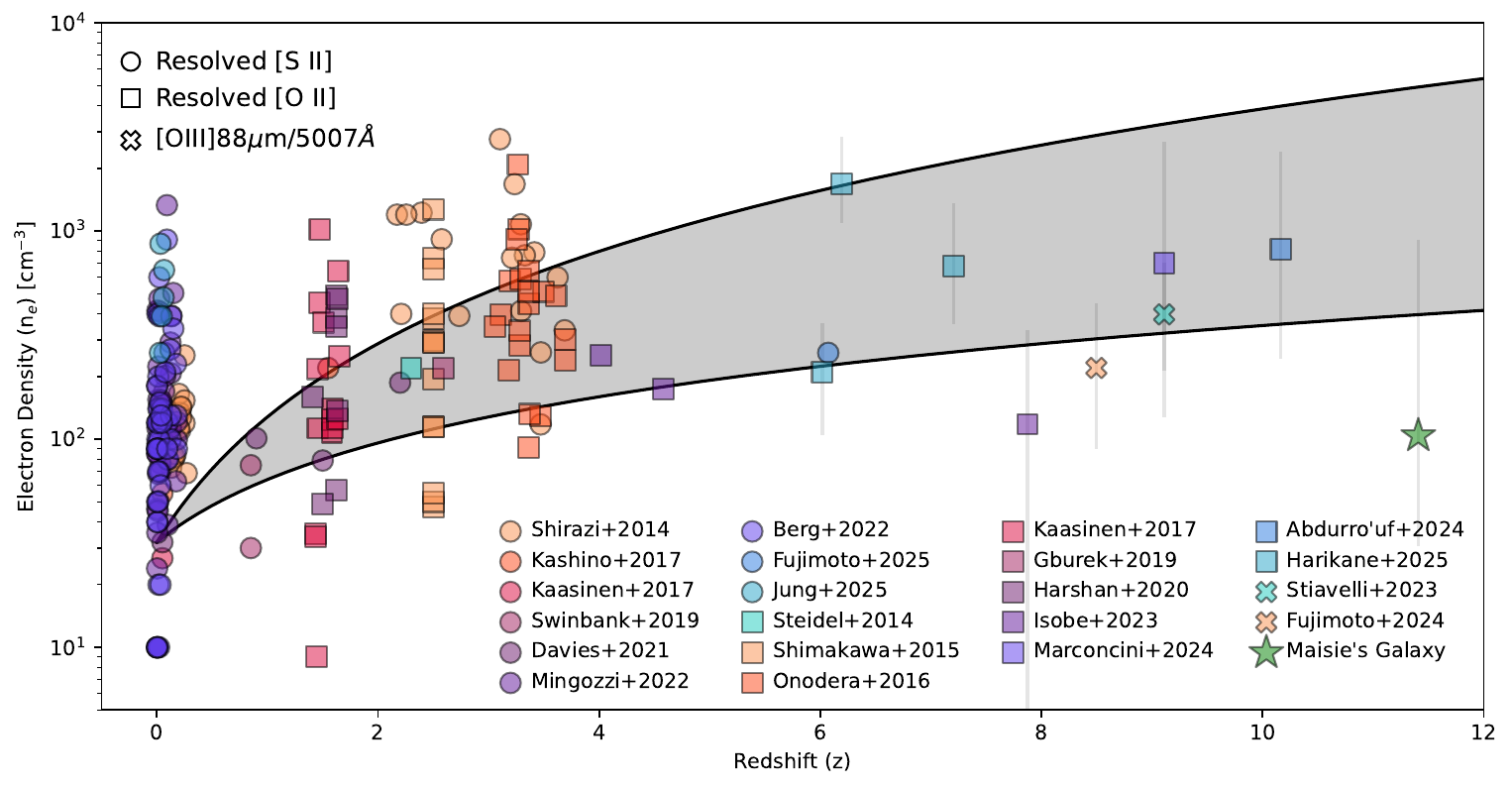}
    \caption{Measured electron densities ($n_e$) vs redshift ($z$) for sources from the literature and including the new measurement for Maisie's Galaxy (green star). Those with $n_e$ derived from resolved \sii\ emission line ratios are plotted with triangles, those from resolved \oii\ emission line ratios are plotted with squares, those from resolved \ciii\ ratios with circles, and those from the \oiii$\lambda 88\mu$m/5007\AA\ ratio with an x. We also shade the region between the curves of $n_e \propto (1+z)$ and $n_e \propto (1+z)^2$ as previous studies hinted at a redshift evolution fit by a power law in this range \citep{davies21, isobe23}. The measurement for Maisie's Galaxy is consistent with this trend within errors.}
    \label{fig:nevz}
\end{figure*}

Our coadded spectrum yields a line ratio measurement of $R_{\oii} = \oii \lambda 3729/3727$\AA\ = $1.35 \pm 0.39$, which can be used to place constraints on the electron density ($n_e$) of Maisie's Galaxy \citep{berg21}. The critical densities ($n_c$) of these lines are $n_{c,\lambda3727}\sim1.5\times 10^4$ and $n_{c,\lambda3729}\sim3.4\times 10^3$, which means that we are somewhat biased given that our only density tracer is 1: low ionization (13.6eV) and 2: low critical density (if there is high density gas, we are not sensitive to it with \oii\ alone). To measure this, we use the photoionization models generated by \citet{kewley19a}, which give the relationship between $R_{\oii}$ and $n_e$ as a function of electron temperature $T_e$ in equations B1 and B2 of that paper. In Figure \ref{fig:te}, the curves for different $\log(T_e)$, ranging from 3.375 to 4.625, are shown in various colors. The black dashed line represents the model from \citet{sanders16}, which aligns with the \citet{kewley19a} model for $T_e = 10,000$ K. The vertical gray dashed lines mark the range of electron densities for which $R_{\oii}$ is a useful density diagnostic \citep{kewley19a}. As we cannot directly measure the $T_e$ for Maisie's Galaxy, we assume the same as that measured for MACS0647-JD1, a $z=10.2$ lensed galaxy, of $T_e = 17,000$ K \citep{hsiao24b, abdurrouf24}. Given our measured $R_{\oii}$ and error we infer $n_e = 77.39^{+503.09}_{-77.39}$. If we assume a standard $T_e = 10,000$ K we get a $n_e = 60.78^{+384.21}_{-60.78}$. Here, we set the lower error bounds to zero to prevent negative values, as these are unphysical. We also note that the error on the $R_{\oii}$ drops the measurement below the dashed lines in Figure \ref{fig:te}, past the point at which it becomes a useful diagnostic for $n_e$. 

As a secondary check for our measurements, we also used the code \textsc{PyNeb} \citep{pyneb} to determine the $n_e$ from our observed $R_{\oii}$. Here, we used the same two values for $T_e$ and ran an MCMC chain with 1000 iterations to obtain the $1\sigma$ errors (16th and 84th percentiles). Using this method, we find that the $n_e = 108.56^{+873.90}_{-35.37}$ when $T_e = 17,000$ K and $n_e = 96.79^{+664.00}_{-32.95}$ when $T_e = 10,000$ K. In general, these two methods agree within error that the electron density is somewhere in the range of $n_e \sim  10^{2-3}$ cm$^{-3}$. For the following analysis, we use the value for $n_e$ obtained from the \textsc{PyNeb} code when using a $T_e = 17,000$ K, the same as that measured for the closest source in redshift with a direct measurement, MACS0647-JD1 at $z = 10.2$ \citep{abdurrouf24}.

We put this source in the context of the other galaxies with $n_e$ measurements in Figure \ref{fig:nevz}. Here we plot galaxies that have $n_e$ from resolved measurements of the \sii\ $\lambda6716/\lambda6731 $ doublet with circles \citep{shirazi14, kashino17, kaasinen17, swinbank19, davies21, mingozzi22, berg22, fujimoto25, jung25}, galaxies with resolved $R_{\oii}$ as squares \citep{steidel14, shimakawa15, onodera16, kaasinen17, gburek19, harshan20, isobe23, marconcini24, abdurrouf24, harikane25}, %galaxies from resolved \ciii\ $\lambda1907/\lambda1909$ with circles \citep{maseda17, topping25},
and galaxies from the \oiii\ $\lambda88\mu$m/$\lambda5007$\AA\ ratio as x's \citep{fujimoto24, stiavelli23}. We plot Maisie's Galaxy on this figure using a green star. 

We also plot the function form of the curves for $n_e \propto (1+z)$ and $n_e \propto (1+z)^2$ with an anchor of $n_e=32$cm$^{-2}$ at $z=0$ and shade the region in between, as previous studies hinted at a redshift evolution fit by a power law in this range \citep{davies21, isobe23}. While this source and many others at $z>8$ appear to fall below the proposed relation, they all agree within the larger errors. % We also note that the \ciii\ diagnostic can probe a different ionization region of the galaxy than the other diagnostics, which is why these points appear elevated. 

\subsection{Metallicity} \label{sec:metallicity}

The redshift of Maisie's Galaxy is high enough such that the \oiii\ and \hb\ lines are shifted out of the NIRSpec G395M coverage. As we have MIRI/LRS observations that detect the \oiii\ doublet, we can use the \oiii/\oii\ (O32) ratio to measure the metallicity of this galaxy. Using the relation derived from \citet{sanders25c}, we can calculate the metallicity as $$ \mathrm{ log (O32) = 0.697 - 1.245x - 0.869x^2}$$ where x = 12 + log(O/H) - 8.0 and find this relationship is valid for values of Z = 12 + log (O/H) between 7.3 and 8.6 (with Z/Z$_\odot$ = 8.69 ). For Maisie's Galaxy, we calculate a log (O32) $=0.724 \pm 0.191$ which gives us a 12 + log (O/H) = 7.98 $\pm$ 0.16 or Z/Z$_\odot$ = 0.19 $\pm$ 0.08.

Often, the O32 ratio is used to determine metallicity, but when \oiii\ emission is absent due to lack of MIRI coverage at these redshifts, we can use alternative line ratios. Using more than 100 high-redshift sources, \citet{sanders25c} determine the relationship between the \neiii/\oii\ (Ne3O2) ratio and metallicity as $$ \mathrm{ log (Ne3O2) = -0.333 - 1.459x - 1.127x^2}$$ where x = 12 + log(O/H) - 8.0 and find this relationship is valid for values of Z = 12 + log (O/H) between 7.4 and 8.6 (with Z/Z$_\odot$ = 8.69 ). Using our measured line ratio, log (Ne3O2) = -0.219$\pm$ 0.145, we infer a metallicity of 12 + log (O/H) = 7.92 $\pm$ 0.12 or Z/Z$_\odot$ = 0.17 $\pm$ 0.05. This relationship is also discussed in \citet{bian18,jones15,maiolino08}. Using those equations, we infer metallicities that agree with the value reported above. Our SED fitting of this source yielded Z/Z$_\odot$ = 0.13$^{+0.08}_{-0.04}$ for this galaxy, but using these line diagnostics yields a tighter constraint on this value.

\begin{figure}[h!]
    \centering
    \includegraphics[width=0.99\linewidth]{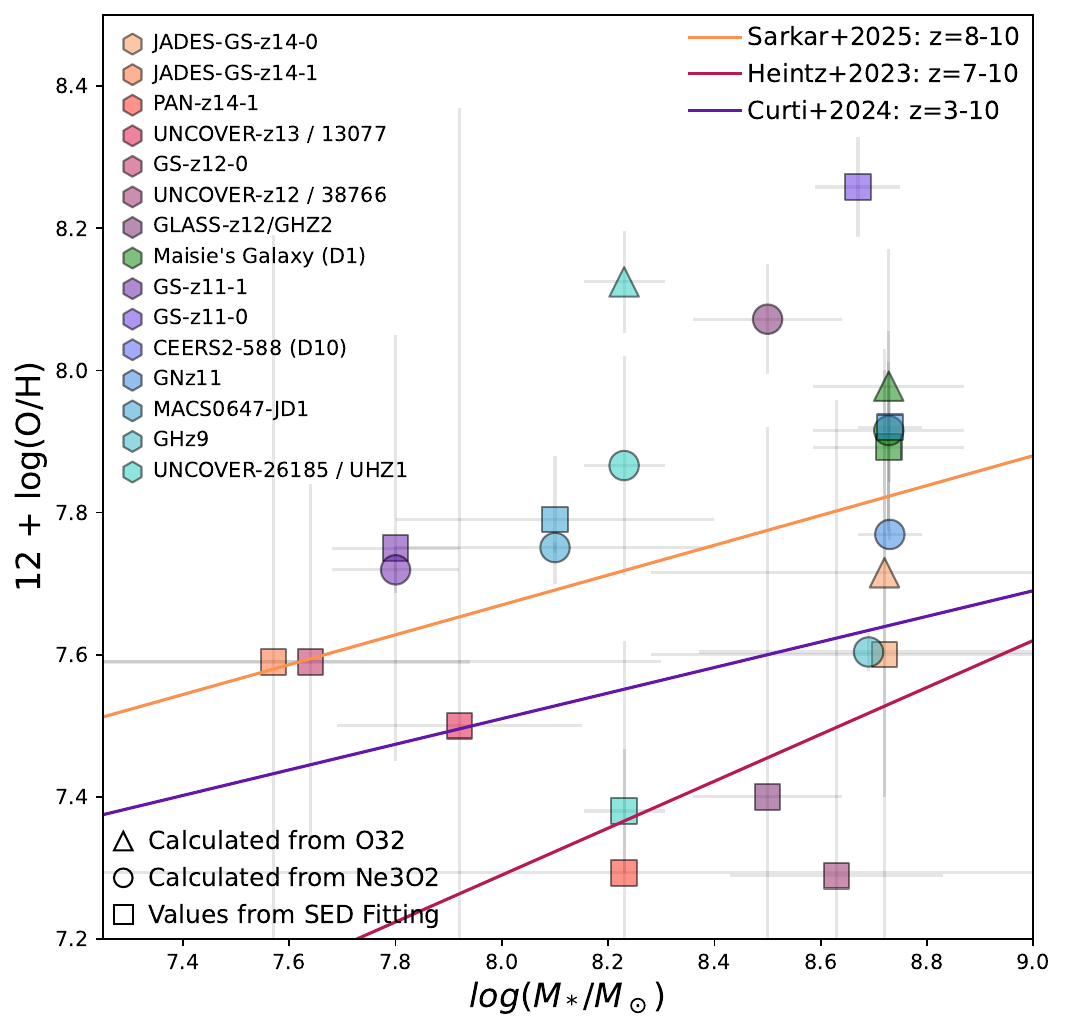}
    \caption{The mass-metallicity relation (MZR) plot with circle points showing sources in the literature with measured \neiii\ and \oii\ emission lines, which have been converted to a metallicity, 12 + log(O/H), using the Ne3O2 ratio. Triangles are those sources with measured \oiii\ lines and metallicity calculated from the O32 ratio. Square points are values from the literature, mostly obtained from SED Fitting. The orange, purple, and red lines are the relations from \citet{sarkar25, heintz23, curti24}, drawn from high-redshift source samples. Stellar masses, log (M$_*$/$M_\odot$), for all sources are derived from SED fitting. While most of our measured metallicities lie above these trend lines, more measurements of this type are required to determine whether these relations change at $z>10$.  }
    \label{fig:mzr}
\end{figure}

To place Maisie's Galaxy in context, while removing as many systematic differences as possible, we performed this same calculation for the 6 other sources in the literature with reported \neiii\ and \oii\ line fluxes at $z>10$ \citep{castellano24, scholtz26, bunker23, hsiao24a, napolitano25a, goulding23} to re-derive the metallicity measurements. We present the metallicities we measure in Table \ref{tab:measurements}, in the Ne3O2 column under 12+log(O/H). We used the O32 ratio for the 3 other sources in the literature with measured \oiii\ emission \citep{hsiao24b,zavala25,Schouws.2025, Alvarez-Marquez.2025,Alvarez-Marquez.2026}, but find that three (GLASS-z12/GHZ2, GNz11, and MACS0647-JD1) have metallicities too low and are thus outside the parameter space. We plot the values from the O32-derived metallicities as triangles in Figure \ref{fig:mzr}, and show them in Table \ref{tab:measurements}. We also compare these values to those reported in the literature for $z>10$ from SED fitting, shown in the SED column under the 12+log(O/H) column in Table \ref{tab:litsources} \citep{carniani24, donnan26, scholtz26, hsiao24b}. We note that metallicity levels of $\sim1-10$\% of the solar value are consistent with theoretical predictions for chemical enrichment in the first galaxies \citep[e.g.,][]{pallottini14,jaacks19,storck2026}, with Maisie's Galaxy representing a more advanced stage of chemical evolution.

This source has a stellar mass, log (M$_*$/M$_\odot$) = 8.14$^{+0.23}_{-0.21}$, as measured from our SED fitting (see \S\ref{sec:bagpipes}). Using mass values from the literature, derived from various SED-fitting methods for sources with confirmed Ne3O2 and/or O32 measurements, we can place our sources on the mass-metallicity relation (MZR) plot (See Figure \ref{fig:mzr}). Here, we plot the points with our calculated 12 + log(O/H) values from Ne3O2 as circles and from O32 as triangles. Reported values in the literature for 12+log(O/H) for sources at $z>10$, which were mostly derived from SED fitting, are shown as squares \citep{carniani24, donnan26, curtislake23, scholtz26, harikane26, hsiao24a} and values are listed in Table \ref{tab:measurements}. The three values for Maisie's Galaxy are plotted in green, and all agree within error. We also show the high-redshift relations derived from \citet{sarkar25,heintz23,curti24} as orange, purple, and red lines, respectively, but note that these samples are limited to $z<10$, whereas our sample is entirely at $z>10$.

\subsection{Ionization Parameter} \label{sec:logu}

\begin{figure}
    \centering
    \includegraphics[width=0.99\linewidth]{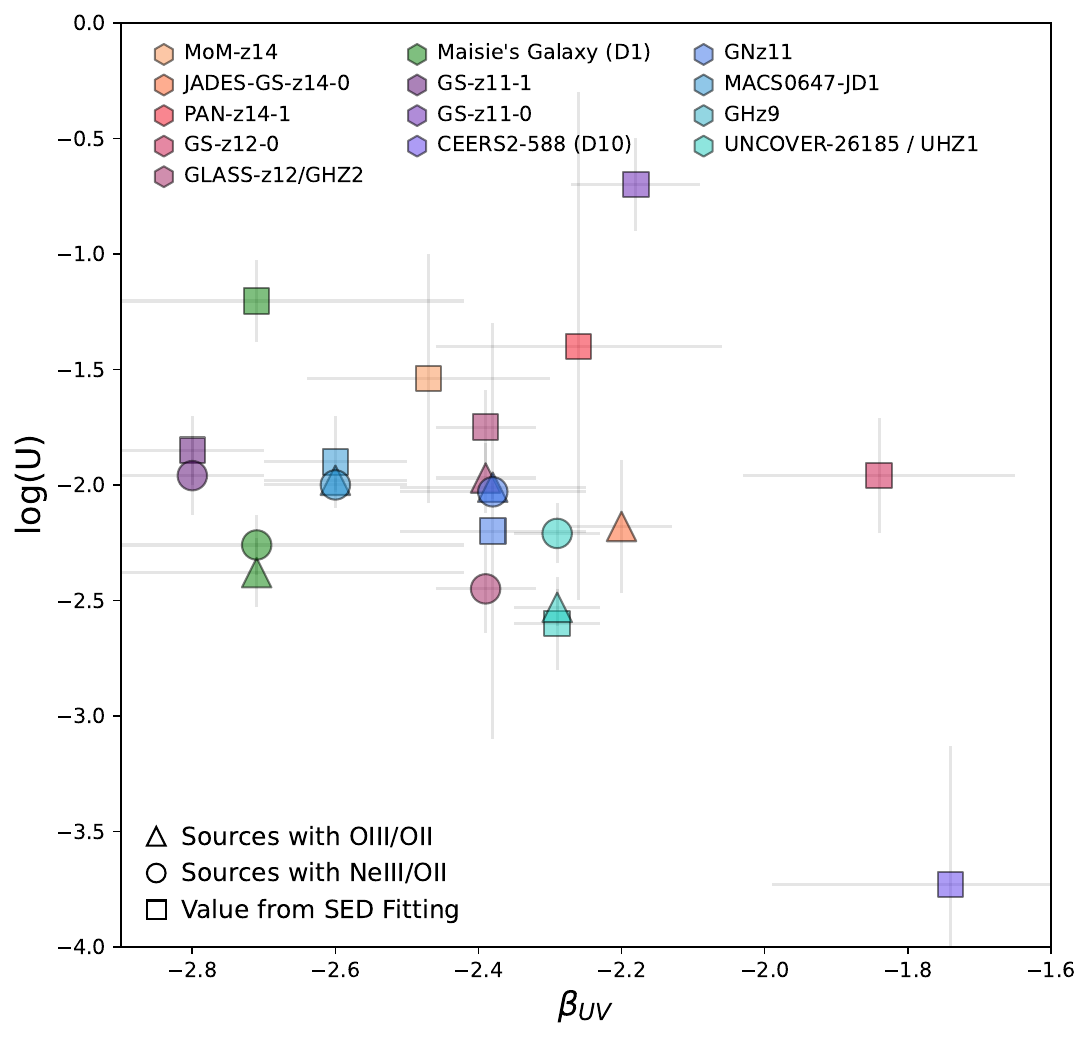}
    \caption{$\log$(U) vs $\beta_{UV}$ for sources with reported values in the literature (squares, right) and sources with reported \oii, \neiii, and \oiii\ line fluxes (left) which were calculated using the equations from \citet{witstok21, kewley19a}. Sources with measured Ne3O2 ratios are plotted with circles, and those with O32 are plotted with triangles. Maisie's galaxy is plotted in green in both panels. Line-ratio determined $\log$(U) values do not seem to trend with $\beta_{UV}$, and are largely lower than those inferred from SED fitting.}
    \label{fig:betauvlogu}
\end{figure}

The measured $\log$(Ne3O2) $= -0.219 \pm 0.145$ ratio for Maisie's Galaxy can be used to constrain the ionization parameter, $\log$(U) \citep{levesque14}. By using a sample of high-redshift galaxies from \citet{vanzella20} and high-redshift analog 'green pea' galaxies from \citet{jaskot13} as well as a large sample of SDSS galaxies, \citet{witstok21} fit the relation between the Ne3O2 ratio and the O32 ratio. They find these ratios to be strongly correlated at high O32 values with $$\log_{10}(Ne3O2) = 0.9051 \log_{10}(O32)-1.078.$$ Thus, the Ne3O2 ratio can be used to calculate the ionization parameter instead of (or in addition to) the traditional O32 ratio. This allows for measuring this diagnostic beyond those derived from single-star photoionization models \citep{diaz00} where $$\log_{10}(U) = 0.80 \log_{10} (O32) - 3.02.$$ Thus, we can calculate the ionization parameter with Ne3O2 as: $$\log_{10}(U) = 0.884 \log_{10}(Ne3O2)-2.07$$ for those sources that fall in the high O32 regime with high $\log$(U), as expected for high-redshift sources \citep[e.g.][]{stark17}.

We use the equation for the Ne3O2 ratio for Maisie's Galaxy and find that $\log$(U)$= -2.26\pm0.09$, and when using the O32 ratio we calculate a simliar value of $\log$(U)$= -2.38\pm0.15$. We perform this same calculation for all the sources in the literature that have reported \neiii\ and \oii\ line fluxes  \citep{castellano24,scholtz26, bunker23, abdurrouf24, napolitano25a, goulding23}. For the 5 sources in the literature that have \jwst/MIRI spectroscopic observations and thus have an \oiii\ line flux measurement in addition to Maisie's Galaxy, we compute the $\log$(U) value using the O32 ratio and the equation above, and plot them as triangles in Figure \ref{fig:betauvlogu} \citep{Helton.2025a, zavala25, hsiao24b}. For one of those sources that has both measurements, MACS0647-JD1, the $\log$(U) measurements agree when using these equations. For the other source with all three emission lines, GLASS-z12/GHZ2, the resulting $\log$(U) values do not agree but are both lower than the value obtained from SED fitting \citep{castellano24}. Values calculated from the line ratios (Ne3O2 and/or O32) are reported in Table \ref{tab:measurements} and those reported in the literature (SED) are listed in Table \ref{tab:litsources}.

We plot the $\log$(U) values calculated from the Ne3O2 ratio as circles in Figure \ref{fig:betauvlogu}, with Maisie's in green, and those from the O32 ratio as triangles as a function of UV slope, $\beta_{UV}$. We also plot the values obtained from SED fitting and reported in the literature as squares on the right side of Figure \ref{fig:betauvlogu}, with Maisie's Galaxy in green. We note that the range of log(U) values from SED fitting is much larger and seems to correlate with $\beta_{UV}$, while those measured from the line ratios all fall within the $\log$(U) $\sim 2-2.5$ range, with no apparent dependence on $\beta_{UV}$. In most cases, the line-ratio-derived log(U) is lower than that measured by SED fitting methods. This implies that either SED fitting processes are not accurately accounting for line fluxes and ratios, or that we need more diagnostic data for galaxies at these redshifts to inform our modeling. 

% Using equation 1 from their paper: $R = k_0 + k_1x + k_2x^2 + k_3x^3$, where R = $\log(\neiii/\oii)$ = Ne3O2, $x=\log_{10}(q $ [cm s$^{-1}$]), and the coefficients are metallicity dependent such that for $Z = [0.001, 0.004, 0.008, 0.02]$ (or Z = [5\%$Z_{\odot}$, 20\%$Z_{\odot}$, 40\%$Z_{\odot}$, $Z_{\odot}$]) we have $k_0$ = [56.3416, 53.5278, 48.7182, 24.7917] , $k_1$ = [-23.1202, -22.2764, -20.6263, -11.7739], $k_2$ = [3.0064, 2.93137, 2.7508, 1.66339], and $k_3$ = [-0.124519, -0.122954, -0.116949, -0.0732419] \citep{levesque14}. We solve this equation for the roots and find that the ionization parameter, $\log(\mathrm{U}) = q/c$, for this range of metallicities is $\log(\mathrm{U}) = [-2.65 \pm 0.09, -2.59 \pm 0.11, -2.51 \pm 0.13, -2.07 \pm 0.22]$ for Maisie's Galaxy. 

% Our measurement for MACS0647–JD is similar to the
% value log(U) = −1.8 ± 0.3 measured for GHZ2/GLASSz12 at z = 12.33 (Castellano et al. 2024; Zavala et al.
% 2024) and log(U) = −2.2 ± 0.9 measured for GN-z11 at
% z = 10.60 (Bunker et al. 2023). These and other similar
% measurements of log(U) ∼ −2 for high-redshift galaxies
% reveal strong ionization parameters, comparable to the
% more starbursting environments we see at low-redshift,
% but overall higher than the non-star forming galaxies
% (e.g., Kewley & Dopita 2002; Mingozzi et al. 2024).

% \begin{table*}[]
%     \centering
%     \begin{tabular}{c|c|c|c|c|c|c|c|c}

\begin{deluxetable*}{r@{\hspace{0.1cm}}|c@{\hspace{0.1cm}}|c@{\hspace{0.1cm}}c@{\hspace{0.1cm}}c@{\hspace{0.1cm}}|c@{\hspace{0.1cm}}c@{\hspace{0.1cm}}c@{\hspace{0.1cm}}c@{\hspace{0.1cm}}|l@{\hspace{0.1cm}}}[!ht] \label{tab:litsources}
\tablecaption{\label{table:litsources}Spectroscopically-Measured \& Photometrically-Derived Physical Conditions of Galaxies at $z>10$ from the Literature}
\tabletypesize{\footnotesize} % this is the command for fontsize
\tablecolumns{10}
\tablehead{ \colhead{\multirow{2}{*}{Source Name}}  & \colhead{\multirow{2}{*}{Redshift}} & \multicolumn{3}{c}{Line Fluxes [10$^{-19}$ erg/s/cm$^2$] } & \multicolumn{4}{c}{SED-Derived Properties} & \multirow{2}{*}{Ref.} ~\vspace{-1mm} \\
    \cmidrule(lr){3-5} \cmidrule(lr){6-9} ~\vspace{-6mm} \\
    \colhead{} & \colhead{} & \colhead{$f_{\oiii}$} & \colhead{$f_{\oii}$} & \colhead{$f_{\neiii}$}  & \colhead{$\log$(M$_*$)} & \colhead{SFR$_{10}$} & \colhead{12+$\log$(O/H)} & \colhead{$\log$(U)}
    }
      % & & & Ne3O2 & SED & Ne3O2 & O32 & SED & \\
     % \hline
    \startdata 
MoM-z14 & $14.44 \pm 0.02$ & ------ & ------ & ------ & $8.10 \pm0.30$ & $13.00 \pm 3.70$ & ------ & $-1.54 \pm 0.54$ & 37 \\
JADES-GS-z14-0 & $14.1796 \pm 0.0007$ & $23.70 \pm 2.10$ & $2.50 \pm 1.90$ & ------ & $8.72 \pm0.44$ & $9.60 \pm 2.20$ & $7.60 \pm 0.40$ & ------ & 14,25,26,28,33 \\
JADES-GS-z14-1 & $13.86 \pm 0.04$ & ------ & ------ & ------ & $7.57 \pm0.37$ & $2.32 \pm 0.66$ & $7.59 \pm 0.60$ & ------ & 14, 35 \\
PAN-z14-1 & $13.53 \pm 0.06$ & ------ & ------ & ------ & $8.23 \pm1.14$ & $4.80 \pm 13.60$ & $7.29 \pm 0.33$ & $-1.40 \pm 1.10$ & 36 \\
UNCOVER-z13 / 13077 & $13.079 \pm 0.01$ & ------ & ------ & ------ & $7.92 \pm0.23$ & $1.95 \pm 0.41$ & $7.50 \pm 0.87$ & ------ & 10,20,29 \\
GS-z12-0 & $12.48 \pm 0.5$ & ------ & ------ & ------ & $7.64 \pm0.66$ & $1.62 \pm 0.26$ & $7.59 \pm 0.25$ & $-1.96 \pm 0.25$ & 2,15,16 \\
UNCOVER-z12/38766 & $12.393 \pm 0.004$ & ------ & ------ & ------ & $8.63 \pm0.20$ & $1.40 \pm 2.81$ & $7.29 \pm 0.67$ & ------ & 9,10,20,29 \\
GLASS-z12/GHZ2 & $12.34 \pm 0.04$ & $47.00 \pm 5.00$ & $2.70 \pm 1.10$ & $6.40 \pm 0.80$ & $8.50 \pm0.14$ & $7.00 \pm 2.00$ & $7.40 \pm 0.52$ & $-1.75 \pm 0.16$ & 13,18,19,31,34 \\
Maisie's Galaxy (D1) & $11.408 \pm 0.006$ & $10.10 \pm 3.44$ & $1.90 \pm 0.53$ & $1.15 \pm 0.21$ & $8.73 \pm0.14$ & $3.99 \pm 0.49$ & $7.89 \pm 0.12$ & $-1.20 \pm 0.18$ & 1,6,29 \\
GS-z11-1 & $11.272 \pm 0.0028$ & ------ & $0.56 \pm 0.15$ & $0.75 \pm 0.15$ & $7.80 \pm0.12$ & $2.09 \pm 0.53$ & $7.75 \pm 0.30$ & $-1.85 \pm 0.15$ & 39 \\
GS-z11-0 & $11.1221 \pm 0.0006$ & ------ & ------ & ------ & $8.67 \pm0.08$ & $1.70 \pm 1.20$ & $8.26 \pm 0.07$ & $-0.70 \pm 0.20$ & 2,21,24 \\
CEERS2-588 (D10) & $11.04 \pm 0.5$ & ------ & $4.52 \pm 0.59$ & ------ & $9.10 \pm0.10$ & $8.20$ & $8.59 \pm 0.21$ & $-3.73 \pm 0.60$ & 12,38 \\
CAPERS\_UDS\_z11 & $11.013 \pm 0.028$ & ------ & $3.72 \pm 0.73$ & ------ & $8.70 \pm0.20$ & $43.00 \pm 10.00$ & ------ & ------ & 27 \\
GNz11 & $10.6034 \pm 0.0013$ & $136.00 \pm 14.00$ & $8.90 \pm 0.80$ & $10.00 \pm 0.80$ & $8.73 \pm0.06$ & ------ & $7.92 \pm 0.08$ & $-2.20 \pm 0.90$ & 8,23,32 \\
MACS0647-JD1 & $10.1674 \pm 0.0002$ & $226.00 \pm 21.00$ & $10.80 \pm 1.40$ & $15.70 \pm 1.00$ & $8.10 \pm0.30$ & $5.00 \pm 0.60$ & $7.79 \pm 0.09$ & $-1.90 \pm 0.20$ & 4,11,17 \\
GHz9 & $10.145 \pm 0.01$ & ------ & $1.99 \pm 0.39$ & $4.11 \pm 0.41$ & $8.69 \pm0.32$ & ------ & ------ & ------ & 22,30 \\
UNCOVER-26185/UHZ1 & $10.073 \pm 0.002$ & $15.80 \pm 1.70$ & $4.67 \pm 0.91$ & $3.24 \pm 0.91$ & $8.23 \pm0.08$ & $1.70 \pm 0.40$ & $7.38 \pm 0.09$ & $-2.60 \pm 0.20$ & 3,5,7,20,40 \\
    \enddata 
\vspace{2mm}
\tablecomments{Data for sources in the literature with reported \oii\ and/or \neiii\ or \oiii\ line fluxes at $z>10$, or with relevant physical properties derived from SED fitting. Columns are: Source Name(s); Redshift; emission line fluxes in 10$^{-19}$ erg s$^{-1}$ cm$^{-2}$ for \oiii$\lambda 5008$, \oii$\lambda\lambda 3726,3729$, and \neiii$\lambda 3968$; SED-derived physical properties including stellar mass, $\log$(M$_*$/M$_\odot$), star formation rate in the last 10 Myr, SFR$_{10}$ in M$_\odot$ yr$^{-1}$, metallicity or 12 + log(O/H), and ionization parameter or $\log$(U); and applicable references.   \\
~\\
References, in chronological order:
1\,=\,\citealt{finkelstein22c}, 
2\,=\,\citealt{curtislake23}, 
3\,=\,\citealt{Castellano.2023}, 
4\,=\,\citealt{Hsiao.2023}, 
5\,=\,\citealt{Roberts-Borsani.2023}, 
6\,=\,\citealt{arrabalharo23}, 
7\,=\,\citealt{goulding23}, 
8\,=\,\citealt{bunker23}, 
9\,=\,\citealt{atek23b},
10\,=\,\citealt{wang23},
11\,=\,\citealt{abdurrouf24}, 
12\,=\,\citealt{ArrabalHaro.2023}
13\,=\,\citealt{castellano24}, 
14\,=\,\citealt{carniani24}, 
15\,=\,\citealt{D'Eugenio.2024}, 
16\,=\,\citealt{Bunker.2024}, 
17\,=\,\citealt{hsiao24b}, 
18\,=\,\citealt{Calabro.2024}, 
19\,=\,\citealt{Zavala.2024}, 
20\,=\,\citealt{Fujimoto.2024}, 
21\,=\,\citealt{Hainline.2024}, 
22\,=\,\citealt{napolitano25a}, 
23\,=\,\citealt{Alvarez-Marquez.2025}, 
24\,=\,\citealt{witstok25}, 
25\,=\,\citealt{Carniani.2025}, 
26\,=\,\citealt{Helton.2025a}, 
27\,=\,\citealt{kokorev25}, 
28\,=\,\citealt{Schouws.2025}, 
29\,=\,\citealt{Tang.2025}, 
30\,=\,\citealt{Napolitano.2025}, 
31\,=\,\citealt{ChavezOrtiz.2025}, 
32\,=\,\citealt{CrespoGomez.2025}, 
33\,=\,\citealt{helton25}, 
34\,=\,\citealt{zavala25},
35\,=\,\citealt{wu25},
36\,=\,\citealt{donnan26}, 
37\,=\,\citealt{Naidu.2026}, 
38\,=\,\citealt{harikane26}, 
39\,=\,\citealt{scholtz26}, 
40\,=\,\citealt{Alvarez-Marquez.2026}.
}
% \tablenotetext{a}{wgfqegwg}
\end{deluxetable*}

% \end{tabular}
% \caption{Caption}
% \label{tab:litsources}
% \end{table*}

% used this:   " = \citealt{%H.%Y}, "
% to export the citations in chronological order from this
% ADS library:
% https://ui.adsabs.harvard.edu/public-libraries/nvc9pzvnR3iemZBYyRYeTA
% NOTE FOR REBECCA -- I pasted these at the top of the bib file, didn't check for duplicates with anything you already had cited in text using a different bib entry lol.

\begin{deluxetable*}{r|cc|c|cc|cc}[!ht] \label{tab:measurements}
\tablecaption{Emission-Line Derived Physical Properties of $z>10$ Galaxies in the Literature}
\tabletypesize{\footnotesize} % this is the command for fontsize
\tablecolumns{8}
\tablehead{ \colhead{\multirow{2}{*}{Source Name}}  & \multicolumn{2}{c}{Line Ratios } & \colhead{\multirow{2}{*}{SFR$_{\oii}$}} & \multicolumn{2}{c}{12 + log(O/H)} & \multicolumn{2}{c}{$\log$(U)} ~\vspace{-1mm} \\
    \cmidrule(lr){2-3} \cmidrule(lr){5-6} \cmidrule(lr){7-8} ~\vspace{-6mm} \\
    \colhead{} & \colhead{log(Ne3O2)} & \colhead{log(O32)} & \colhead{} & \colhead{Ne3O2} & \colhead{O32}  & \colhead{Ne3O2} & \colhead{O32} 
    }
      % & & & Ne3O2 & SED & Ne3O2 & O32 & SED & \\
     % \hline
    \startdata 
%MoM-z14 & ------ & ------ & ------ & ------ & ------ & ------ & ------ \\ 
JADES-GS-z14-0 & ------ & $0.977 \pm 0.367$ & $2.64 \pm 2.11$ & ------ & $7.72 \pm 0.32$ & ------ & $-2.18 \pm 0.29$ \\ 
%JADES-GS-z14-1 & ------ & ------ & ------ & ------ & ------ & ------ & ------ \\ 
%PAN-z14-1 & ------ & ------ & ------ & ------ & ------ & ------ & ------ \\ 
%UNCOVER-z13 / 13077 & ------ & ------ & ------ & ------ & ------ & ------ & ------ \\ 
%GS-z12-0 & ------ & ------ & ------ & ------ & ------ & ------ & ------ \\ 
%UNCOVER-z12 / 38766 & ------ & ------ & ------ & ------ & ------ & ------ & ------ \\ 
GLASS-z12/GHZ2 & $-0.432 \pm 0.211$ & $1.241 \pm 0.183$ & $2.08 \pm 1.00$ & $8.07 \pm 0.08$ & ------ & $-2.45 \pm 0.19$ & $-1.97 \pm 0.15$ \\ 
Maisie's Galaxy (D1) & $-0.219 \pm 0.145$ & $0.724 \pm 0.191$ & $1.30 \pm 0.36$ & $7.92 \pm 0.14$ & $7.98 \pm 0.19$ & $-2.26 \pm 0.13$ & $-2.38 \pm 0.15$ \\ 
GS-z11-1 & $0.127 \pm 0.190$ & ------ & $0.35 \pm 0.13$ & $7.72 \pm 0.03$ & ------ & $-1.96 \pm 0.17$ & ------ \\ 
%GS-z11-0 & ------ & ------ & ------ & ------ & ------ & ------ & ------ \\ 
CEERS2-588 (D10) & ------ & ------ & $2.69 \pm 0.76$ & ------ & ------ & ------ & ------ \\ 
CAPERS\_UDS\_z11 & ------ & ------ & $2.02 \pm 0.64$ & ------ & ------ & ------ & ------ \\ 
GNz11 & $0.049 \pm 0.050$ & $1.184 \pm 0.059$ & $4.88 \pm 1.30$ & $7.77 \pm 0.01$ & ------ & $-2.03 \pm 0.04$ & $-2.01 \pm 0.05$ \\ 
MACS0647-JD1 & $0.079 \pm 0.038$ & $1.230 \pm 0.051$ & $5.38 \pm 1.52$ & $7.75 \pm 0.01$ & ------ & $-2.00 \pm 0.03$ & $-1.98 \pm 0.04$ \\ 
GHz9 & $0.315 \pm 0.197$ & ------ & $0.99 \pm 0.31$ & $7.60 \pm 0.03$ & ------ & $-1.79 \pm 0.17$ & ------ \\ 
UNCOVER-26185/UHZ1 & $-0.159 \pm 0.148$ & $0.529 \pm 0.097$ & $2.39 \pm 0.46$ & $7.87 \pm 0.15$ & $8.12 \pm 0.07$ & $-2.21 \pm 0.13$ & $-2.53 \pm 0.08$ \\ 
    \enddata 
\vspace{2mm}
\tablecomments{Measurements for Maisie's Galaxy and the other sources in the literature with reported \oii\ and/or \neiii\ or \oiii\ line fluxes at $z>10$. Columns are Source Name(s); Line Ratios for $\log$(Ne3O2) and $\log$(O32); SFR$_{\oii}$, which we calculate as described in \S\ref{sec:sfroii}; Metallicity or 12 + log(O/H) as calculated using the Ne3O2 ratio and/or the O32 ratio and described in \S\ref{sec:metallicity}; Ionization Parameter or $\log$(U) as calculated using the Ne3O2 ratio and/or the O32 ratio and described in \S\ref{sec:logu}. 
}
% \tablenotetext{a}{wgfqegwg}
\end{deluxetable*}

\section{Conclusion}

In this paper, we present combined observations from two \jwst/NIRSpec programs (GO\#5507 and GO \#5943) to achieve $>19$ hour depth on Maisie's Galaxy, an early-identified high-redshift galaxy in the first \jwst\ data \citep{finkelstein22c}. We detect three emission lines, \oii, \neiii, and \hei, and report an updated spectroscopic redshift of $z=11.4082 \pm 0.0057$ for this source, consistent with the value measured from the lower-resolution prism data in \citet{arrabalharo23}. The deep ($>13$ hour) G140M data covers the rest-frame UV of the source, but there are no detections of \lya\ or either Carbon line, unlike in some of the other extreme $z>10$ sources in the literature \citep[e.g.][]{witstok25,bunker23}. We also present the 9-hour MIRI/LRS observation of this source from GO\#3703 and the successful detection of the \oiii\ doublet. 

We use the measured \oii\ emission line feature to determine a star-formation rate of $SFR_{\oii} = 1.29 \pm 0.36$ M$_{\odot}$ yr$^{-1}$, comparatively low for the other $z>10$ sources in the literature with \oii\ line detections. We also find that the SFR calculated from \oii\ is lower than that reported by SED fitting, which could indicate that \oii\ is not a reliable SFR indicator at these epochs, or that our modeling needs improvement. We also find that these $z>10$ galaxies all fall along the star-formation main sequence (SFMS) line as derived for galaxies at $9<z<12$ in the CEERS field \citep{cole25}. More robust line detections from sources at $z>10$ are needed, as well as longer-wavelength \jwst/MIRI observations for \ha\ detections to use as an SFR tracer.

Due to the resolution of our NIRSpec/G395M data and the high redshift of this source, we are able to put some constraints on the doublet ratio of the \oii\ emission line. We measure an \oii$\lambda 3729$/$\lambda 3726 = 1.35 \pm 0.39$, which can provide an estimate of the electron density, $n_e$ of this source. Using the next-closest-in-redshift source with a direct electron temperature, $T_e$, measurement of 17000K, MACS0647-JD1 \citep{abdurrouf24}, we imply an $n_e = 103.91^{+695.79}_{-30.72}$ for Maisie's Galaxy. Putting this in the context of other measurements as a function of redshift, we find that given the high errors, it is consistent with the notion that $n_e \propto (1+z)$ to $(1+z)^{2}$ \citep{davies21, isobe23}. This is the highest-redshift source with a density measurement, so more data at these depths and higher resolution are needed for $z>10$ galaxies to determine how density trends in the early Universe.

Since we detect the \neiii\ emission line, we can use the \neiii/\oii\ (Ne3O2) ratio to determine the galaxy's metallicity \citep{sanders25c}. We measure a log Ne3O2 $= -0.219 \pm 0.145$ for Maisie's galaxy, giving us a metallicity 12 + log(O/H) $=7.92\pm0.12$ or $Z/Z_{\odot}=0.17\pm0.05$. Using the MIRI data, we also measure the \oiii\ line and can use the \oiii/\oii\ (O32) ratio to measure metallicity in a more traditional way. Using this ratio and the equations from \citet{sanders25c}, we derive a 12 + log(O/H) $=7.98\pm0.16$ or $Z/Z_{\odot}=0.19\pm0.08$. These values agree within error, indicating that at these redshifts, when MIRI observations are unavailable, the Ne3O2 ratio can suffice. This metallicity is relatively high compared to other sources with the same measurements at $z>10$, and most of those fall above the relations derived from $z=3-10$ galaxies. Perhaps this galaxy has recently undergone periods of star formation, enriching the gas to higher metallicities than those expected in the early Universe. Or maybe these diagnostics, calibrated for later galaxies, do not apply to these younger sources. More observations detecting lines such as \oiii\ with \jwst/MIRI would be beneficial for calibrating such relations and answering these questions.

Using the Ne3O2 emission line ratio, we can infer the ionization parameter, $\log$(U), for this source based on the relation between Ne3O2 and O32 at high O32, as expected and measured in the early Universe \citep{witstok21}. For Maisie's Galaxy, we find that $\log$(U) $=-2.26\pm0.13$ when calculated using the Ne3O2 ratio, which is lower than that inferred from SED fitting. We compare this to the value we measure using the O32 ratio, $\log$(U) $= -2.38\pm 0.15$, and find that it agrees with that derived from the Ne3O2 ratio. In fact, many of the $\log$(U) measurements calculated using Ne3O2 or O32 for sources in the literature with both lines detected are lower than those obtained from SED fitting. We also find that there is little to no dependence on $\beta_{UV}$ as might be implied by SED fitting routines. Clearly, more data and more robust detections of diagnostic lines at these redshifts are required to determine the properties of these $z>10$ galaxies.

The status of Maisie's Galaxy as one of the first ``normal'' galaxies, with a moderate SFR and metallicity, implies an already advanced evolutionary state, pushing its assembly history to even higher redshifts. A related question then is: How could such a relatively mature galaxy have emerged so early in cosmic history? Approximately, one can estimate that 100\,Myr were needed to produce the $\sim$$10^8\,M_{\odot}$ in stars, and $\sim$$10^5\,M_{\odot}$ in metals, assuming standard nucleosynthetic yields. This would push the true birth of Maisie's Galaxy to $z\sim 14.5$, allowing a multi-generational cycle of star formation and supernova-driven chemical enrichment that was able to establish a high degree of complexity already at the very beginning of galaxy formation \citep[e.g.,][]{Jeon2019,Yajima2023}. A more complete census of the mix of first-galaxy types is needed to fully assess such questions of ultimate galactic origins, but these types are now clearly within reach.

Maisie's Galaxy is a more typical galaxy in the early Universe, lacking extreme emission-line detections or inferred physical properties. It is imperative to go deeper with our observations if we want to study the more typical population of galaxies in this epoch. \jwst/MIRI observations to detect the \oiii\ and \ha\ emission lines are the best next step to understanding the physical properties of galaxies in the first 500\,Myr.

%% Please use the acknowledgment and contribution environments. This will 
%% be anonomyized when the "anonymous" style option is used. 
\begin{acknowledgments}

This work is based on observations made with the NASA/ESA/CSA \emph{JWST}. The data were obtained from the Mikulski Archive for Space Telescopes at the Space Telescope Science Institute, which is operated by the Association of Universities for Research in Astronomy, Inc., under NASA contract NAS 5-03127 for \JWST. These observations are associated with \JWST\ Cycle 3 GO programs 5507 and 5943. Support for the programs JWST-GO-5507 and JWST-GO-5943 was provided by NASA through a grant from the Space Telescope Science Institute (STScI), which is operated by the Association of Universities for Research in Astronomy (AURA), Incorporated, under NASA contract NAS5-26555.
The material is based upon work supported by NASA under award number 80GSFC24M0006. Support to MAST for these data is provided by the NASA Office of Space Science via grant NAG5–7584 and by other grants and contracts.

RLL and AY appreciate support from a Giacconi Fellowship at the Space Telescope Science Institute, which is operated by the Association of Universities for Research in Astronomy, Inc., under NASA contracts NAS 5-26555 and NAS5-03127.
TAH's research is supported by an appointment to the NASA Postdoctoral Program at NASA Goddard Space Flight Center, administered by Oak Ridge Associated Universities under contract with NASA, as well as by the University of Maryland, Baltimore County, and the Center for Space Sciences and Technology.
J.A.-M. acknowledges support by grants PID2024-158856NA-I00 and PID2021-127718NB-100 from the Spanish Ministry of Science and Innovation/State Agency of Research MCIN/AEI/10.13039/501100011033 and by “ERDF A way of making Europe”. 

\end{acknowledgments}

\begin{contribution}
%%This section gives authors the space to recognize author contributions. The text inside this environment is NOT counted towards the total word quanta. 

RLL is Co-PI of the THRILS program, planned the observations, did the analysis, and wrote the manuscript.
TAH is Co-PI of the THRILS and C3PO programs, planned the observations, assisted with the plotting software, and edited the manuscript.
SLF initially identified the source, developed the research concept, provided scientific expertise, and edited the manuscript.
PAH reduced the data from both NIRSpec programs using methods described in this work. 
CJP and WH are also Co-PI's of the C3PO program, provided scientific expertise, and edited the manuscript. CJP also ran the BAGPIPES SED fitting for this paper.
JA-M reduced the data from the MIRI/LRS program.
RL measured the emission features in the MIRI data.
JAZ was the PI of the MIRI/LRS program.

All other authors provided scientific discussion during analysis and/or feedback during the writing process.

%%
%% Authors can use the Contributor Role Taxonomy (CRediT) at
%% for ideas on how write a good statement tailored to their needs.

\end{contribution}

\facilities{JWST(NIRSpec)}

\software{astropy \citep{2013A&A...558A..33A,2018AJ....156..123A,2022ApJ...935..167A},  
          %Cloudy \citep{2013RMxAA..49..137F}, 
          %Source Extractor \citep{1996A&AS..117..393B},
          IDL \citep{landsman93},
          \jwst\ Pipeline \citep{bushouse25},
          SAOImage DS9 \citep{SmithsonianAstrophysicalObservatory.2000},
          Source Extractor \citep{1996A&AS..117..393B},
          BAGPIPES \citep{carnall18,bagpipes},
          PyNeb \citep{pyneb}
          }


%\appendix

%% For this sample we use BibTeX plus aasjournalv7.bst to generate the

\bibliography{MaisieOII}{}
\bibliographystyle{aasjournalv7}

\allauthors

~\vspace{1cm}\\
$^*$ Giacconi Postdoctoral Fellow\\
$^\dagger$ NASA Postdoctoral Fellow\\
$^\ddagger$ NSF Graduate Research Fellow\\


\end{document}